\documentstyle[preprint,aps,graphicx,floats,amsmath]{revtex}

\newcommand{\sla}[1]{/\!\!\!#1}

\begin{document}

\preprint{$
\begin{array}{l}
\mbox{UB-HET-04-03}\\[-3mm]
\mbox{Fermilab-Pub-04-344-T}\\[-3.mm]
{\rm November}~2004 \\ [3mm]
\end{array}
$}

\title{Probing Electroweak Top Quark Couplings at Hadron Colliders}

\author{U.~Baur\footnote{baur@ubhex.physics.buffalo.edu\\[-12.mm]}}

\address{Department of Physics,
State University of New York, Buffalo, NY 14260, USA\\[-2.mm]}

\author{A.~Juste\footnote{juste@fnal.gov\\[-12.mm]}}

\address{Fermi National Accelerator Laboratory, Batavia, IL 60510,
USA\\[-2.mm] } 

\author{L.H.~Orr\footnote{orr@pas.rochester.edu\\[-12.mm]} $\;\;$ and $\;\;$ 
        D.~Rainwater\footnote{rain@pas.rochester.edu\\[-12.mm]}}

\address{Department of Physics and Astronomy, University of Rochester,
Rochester, NY 14627, USA\\[-2.mm]}

\maketitle

\begin{abstract}
\baselineskip13.pt 
We consider QCD $t\bar{t}\gamma$ and $t\bar{t}Z$ production at hadron
colliders as a tool to measure the $tt\gamma$ and $ttZ$ couplings.
At the Tevatron it may be possible to perform a first, albeit not very
precise, test of the $tt\gamma$ vector and axial vector couplings in
$t\bar{t}\gamma$ production, provided that more than 5~fb$^{-1}$ of
integrated luminosity are accumulated.  The $t\bar{t}Z$ cross section 
at the Tevatron is too small to be observable.  At the CERN Large Hadron
Collider (LHC) it will be possible to probe the $tt\gamma$ couplings
at the few percent level, which approaches the precision which one
hopes to achieve with a next-generation $e^+e^-$ linear collider.  The
LHC's capability of associated QCD $t\bar{t}V$ ($V=\gamma,\,Z$)
production has the added  
advantage that the $tt\gamma$ and $ttZ$ couplings are not entangled.
For an integrated luminosity of 300~fb$^{-1}$, the $ttZ$ vector (axial 
vector) coupling can be determined with an uncertainty of $45-85\%$ 
($15-20\%$), whereas the dimension-five dipole form factors can be 
measured with a precision of $50-55\%$.  The achievable limits improve 
typically by a factor of $2-3$ for the luminosity-upgraded (3~ab$^{-1}$) 
LHC.
\end{abstract}

\newpage


\tightenlines

\section{Introduction}
\label{sec:sec1}

Although the top quark was discovered almost ten years
ago~\cite{topcdf,topd0}, many of its properties are still only poorly
known~\cite{Chakraborty:2003iw}.  In particular, the couplings of the
top quark to the electroweak (EW) gauge bosons have not yet been
directly measured.  The large top quark mass~\cite{topmass} suggests
that it may play a special role in EW symmetry breaking (EWSB).  New
physics connected with EWSB may thus be found first in top quark
precision observables.  A possible signal for new physics are
deviations of the $tt\gamma$, $ttZ$ and $tbW$ couplings from the
values predicted by the Standard Model (SM).  For example, in
technicolor and other models with a strongly coupled Higgs sector,
anomalous top quark couplings may be induced at the $5-10\%$
level~\cite{examples}.

Current data provide only weak constraints on the couplings of the top
quark with the EW gauge bosons, except for the $ttZ$ vector and axial
vector couplings which are rather tightly but indirectly constrained
by LEP data (see Sec.~\ref{sec:sec2c}); and the right-handed $tbW$
coupling, which is severely bound by the observed $b\to s\gamma$
rate~\cite{Larios:1999au}.  In future, the $tbW$ vertex can be probed
in top quark decays to $Wb$~\cite{t2Wb,Dalitz:1998zn,Affolder:2000xb}, 
single top quark production at hadron
colliders~\cite{Cortese:1991fw,Willenbrock:1986cr,Heinson:1996zm,Boos:1999dd},
$e\gamma$ collisions~\cite{jikia}, and top pair production at an
$e^+e^-$ linear
collider~\cite{Boos:1999ca,Grzadkowski:1998bh,Grzadkowski:2000nx}.
The $tt\gamma$ and $ttZ$ couplings can also be tested in $e^+e^-\to
t\bar{t}$~\cite{Grzadkowski:1998bh,Grzadkowski:2000nx,Lin:2001yq,Abe:2001nq,Aguilar-Saavedra:2001rg,Frey:1995ai,Ladinsky:1992vv},
and in $t\bar{t}V$ ($V=\gamma,\, Z$) production at hadron
colliders~\cite{Zhou:1998bh,Baur:2001si}.  Finally, the process 
$\gamma\gamma\to t\bar{t}$ is also sensitive to $tt\gamma$
couplings~\cite{Poulose:1998sd,Djouadi:1995es}.

At an $e^+e^-$ linear collider with $\sqrt{s}=500$~GeV and an
integrated luminosity of $100-200$~fb$^{-1}$ one can hope to measure
the $ttV$ couplings in top pair production with a few-percent
precision~\cite{Abe:2001nq}.  However, the process
$e^+e^-\to\gamma^*/Z\to t\bar{t}$ is sensitive to both $tt\gamma$
and $ttZ$ couplings and significant cancellations between the various
couplings can occur.  At hadron colliders, $t\bar{t}$ production is so
dominated by the QCD processes $q\bar{q}\to g^*\to t\bar{t}$ and
$gg\to t\bar{t}$ that a measurement of the $tt\gamma$ and $ttZ$
couplings via $q\bar{q}\to\gamma^*/Z^*\to t\bar{t}$ is hopeless.
Instead, the $ttV$ couplings can be measured in QCD $t\bar{t}\gamma$
production, radiative top quark decays in $t\bar{t}$ events
($t\bar{t}\to\gamma W^+W^- b\bar{b}$), and QCD $t\bar{t}Z$
production. $t\bar{t}\gamma$ production and radiative top quark decays
are sensitive only to the $tt\gamma$ couplings, whereas $t\bar{t}Z$
production gives information only on the structure of the $ttZ$
vertex.  This obviates having to disentangle potential cancellations
between the different couplings.  In these three processes one can
also hope to separate the dimension-four and -five couplings which
appear in the effective Lagrangian describing the $ttV$ interactions.
Helicity amplitudes of an operator with dimension $n$ in general grow
with energy, $E$, proportional to $E^{n-4}$.  As a result, the shape
of the photon or $Z$ boson transverse momentum distribution differs
considerably for couplings of different dimensionality.

In this paper we consider $t\bar{t}\gamma$ production (including
radiative top quark decays in $t\bar{t}$ events), and $t\bar{t}Z$
production, at the Tevatron and LHC as a tool to measure the $ttV$
couplings.  We first review the couplings definitions, then discuss
existing bounds on them, as well as constraints from $S$-matrix
unitarity (Sec.~\ref{sec:sec2}).  In Secs.~\ref{sec:sec3}
and~\ref{sec:sec4} we present detailed analyses of $t\bar{t}\gamma$
and $t\bar{t}Z$ production, including all relevant backgrounds.  We
derive sensitivity bounds in Sec.~\ref{sec:sec5}, where we also
present a detailed comparison with the limits anticipated at a  
future $e^+e^-$ linear collider.  We summarize in Sec.~\ref{sec:sec6}.

\section{General \boldmath{$ttV$} Couplings}
\label{sec:sec2}

\subsection{Definition}
\label{sec:sec2a}

The most general Lorentz-invariant vertex function describing the
interaction of a neutral vector boson $V$ with two top quarks can be
written in terms of ten form factors~\cite{Hollik:1998vz}, which are
functions of the kinematic invariants.  In the low energy limit,
these correspond to couplings which multiply dimension-four or -five 
operators in an effective Lagrangian, and may be complex.  If $V$ is 
on-shell, or if $V$ couples to effectively massless fermions, the 
number of independent form factors is reduced to eight.  If, in 
addition, both top quarks are on-shell, the number is further reduced 
to four.  In this case, the $ttV$ vertex can be written in the form
\begin{equation}\label{eq:anomvertex}
\Gamma_\mu^{ttV}(k^2,\,q,\,\bar{q}) = -ie \left\{
  \gamma_\mu \, \left( F_{1V}^V(k^2) + \gamma_5F_{1A}^V(k^2) \right)
+ \frac{\sigma_{\mu\nu}}{2m_t}~(q+\bar{q})^\nu 
   \left( iF_{2V}^V(k^2) + \gamma_5F_{2A}^V(k^2) \right)
\right\} \, ,
\end{equation}
where $e$ is the proton charge, 
$m_t$ is the top quark mass, $q~(\bar{q})$ is the outgoing top
(anti-top) quark four-momentum, and $k^2=(q+\bar{q})^2$.  The terms
$F_{1V}^V(0)$ and $F_{1A}^V(0)$ in the low energy limit are the $ttV$ 
vector and axial vector form factors.  The coefficients 
$F_{2V}^\gamma(0)$ and $F_{2A}^\gamma(0)$ are related to the magnetic 
and ($CP$-violating) electric dipole form factors, $g_t$ and 
$d_t^\gamma$ accordingly:
\begin{eqnarray}
F_{2V}^\gamma(0) & = & Q_t \, {g_t-2\over 2} \,, \\
F_{2A}^\gamma(0) & = & {2m_t\over e} \, d_t^\gamma \,,
\end{eqnarray}
where $Q_t=2/3$ is the top quark electric charge. Similar
relations hold for $F_{2V}^Z(0)$, $F_{2A}^Z(0)$, and the weak magnetic
and weak electric dipole moments, $g_t^Z$ and $d_t^Z$.  At tree level
in the SM,
\begin{alignat}{2} \notag
F^{\gamma ,SM}_{1V} & = -{2\over 3} \, , & \qquad
F^{\gamma ,SM}_{1A} & = 0 \, , \\[1.mm]
F^{Z,SM}_{1V} & = -\frac{1}{4\sin\theta_W\cos\theta_W} 
  \left( 1 - \frac{8}{3} \, \sin^2\theta_W \right) , & \qquad
F^{Z,SM}_{1A} & = \frac{1}{4\sin\theta_W\cos\theta_W} \, , \\[2.mm] \notag
F^{\gamma ,SM}_{2V} & = F^{Z,SM}_{2V} = 0 \, , & \qquad
F^{\gamma ,SM}_{2A} & = F^{Z,SM}_{2A} = 0 \, , 
\end{alignat}
where $\theta_W$ is the weak mixing angle.
The one-loop corrections to $F^\gamma_{1V,A}$ vanish for on-shell
photons~\cite{Hollik:1988ii}. The numerically most important radiative
corrections 
to the $ttZ$ vector and axial vector couplings can be taken into account
by replacing the factor $(1-8\sin^2\theta_W/3)$ in $F^{Z,SM}_{1V}$ by
$(1-8\sin^2\theta_{eff}^t/3)$, where $\sin^2\theta_{eff}^t$ is the
effective mixing angle, and by expressing the remaining factors of
$\sin\theta_W$ and $\cos\theta_W$ in $F^{Z,SM}_{1V,A}$ in terms of the
physical $W$ and $Z$ 
masses. Numerically, the one-loop corrections to $F^V_{1V,A}$ are
typically of ${\cal O}(10^{-3} - 10^{-2})$~\cite{Hollik:1988ii}.
The magnetic and weak magnetic dipole form factors $F^V_{2V}$ receive
contributions of the same magnitude~\cite{Bernabeu:1995gs} at the one
loop level in the SM. However, there is no  
such contribution to the electric and weak electric dipole form
factors, $F^V_{2A}$~\cite{Hollik:1998vz}.

In $t\bar{t}V$ production, one of the top quarks coupling to $V$ is
off-shell.  The most general vertex function relevant for $t\bar{t}V$
production thus contains additional couplings, not included in
Eq.~(\ref{eq:anomvertex}). These additional couplings are irrelevant
in $e^+e^-\to t\bar{t}$, where both top quarks are on-shell. Since
most of the existing literature does not discuss them, we ignore these
additional couplings in the following.

In $e^+e^-\to t\bar{t}$ one often uses the following parameterization for
the $ttV$ vertex:
\begin{equation}\label{eq:gordon}
\Gamma_\mu^{ttV}(k^2,\,q,\,\bar{q}) = ie \left\{
  \gamma_\mu \, \left(  \widetilde{F}_{1V}^V(k^2)
                      + \gamma_5\widetilde{F}_{1A}^V(k^2) \right)
+ \frac{(q-\bar{q})_\mu}{2m_t}
    \left(  \widetilde{F}_{2V}^V(k^2)
          + \gamma_5\widetilde F_{2A}^V(k^2) \right)
\right\} .
\end{equation}
Using the Gordon decomposition, it is easy to show that the form
factors $\widetilde F^V_{iV,A}$ and $F^V_{iV,A}$ ($i=1,\,2$) are
related by
\begin{eqnarray}
\label{eq:rel1}
\widetilde F^V_{1V} & = & -\left( F^V_{1V}+F^V_{2V} \right) \, , \\
\widetilde F^V_{2V} & = & \phantom{-} F^V_{2V} \, , \\
\widetilde F^V_{1A} & = & -F^V_{1A} \, , \\
\label{eq:rel4}
\widetilde F^V_{2A} & = & -iF^V_{2A} \, .
\end{eqnarray}
It should be noted that the Gordon decomposition holds only if both
top quarks are on-shell.  Only in this case are the vertex functions
of Eqs.~(\ref{eq:anomvertex}) and~(\ref{eq:gordon}) equivalent.  We
found that for our processes, $t\bar{t}V$ associated production, 
using the Gordon decomposition results in gross Lorentz violations of
the matrix elements. We therefore base our analysis on the form factors in
Eq.~(\ref{eq:anomvertex}) and use Eqs.~(\ref{eq:rel1}--\ref{eq:rel4})
only in Sec.~\ref{sec:sec5}
to compare the limits we obtain
for $F^V_{iV,A}$ at the Tevatron and LHC with those listed in the
literature for $\widetilde F^V_{iV,A}$.

\subsection{Unitarity Constraints}
\label{sec:sec2b}

The parton-level production cross sections of processes such as
$t\bar{t}\to VV$ or $t\bar{t}\to W^+W^-$ with non-SM $ttV$ couplings
manifestly grow with the parton center of mass energy
$\sqrt{\hat{s}}$.  $S$-matrix unitarity restricts the $ttV$ couplings
uniquely to their SM values at asymptotically high
energies~\cite{unitarity}.  This requires that the couplings
$F^V_{iV,A}$ ($i=1,2$) possess a momentum dependence which ensures
that any deviations of the $F^V_{iV,A}(\hat s)$ from their SM values vanish
for $\hat{s}\gg m_t^2$.  The precise $\hat{s}$-dependence of the couplings
is, of course, unknown.  The simplest possible ansatz is to assume a
constant anomalous coupling for $\sqrt{\hat{s}}<\Lambda$ which
abruptly drops to zero at $\sqrt{\hat{s}}=\Lambda$ (step-function)
where the scale $\Lambda$ is related to the scale of the new physics
generating the anomalous couplings.  This ansatz is generally used
when calculating the contributions of non-standard couplings to loop
observables (see Sec.~\ref{sec:sec2c}).  Here, in order to explore how
$S$-matrix unitarity restricts the anomalous $ttV$ couplings, we use
instead a dipole form factor, similar to the well-known nucleon form
factor,
\begin{equation}\label{eq:ff}
\Delta F^V_{iV,A}(k^2) = 
{\Delta F^V_{iV,A}(0) \over ( 1 + k^2/\Lambda_{FF}^{2})^2 } \quad 
(i = 1,2) \, ,
\end{equation}
where 
\begin{equation}
\Delta F^V_{iV,A}(k^2) = F^V_{iV,A}(k^2) - F^{V,SM}_{iV,A},
\end{equation}
and $\Lambda_{FF}$ is the form factor scale which is analogous to the
scale $\Lambda$ discussed above. 

The values $\Delta F^V_{iV,A}(0)$ are constrained by partial wave
unitarity of the amplitudes $t\bar{t}\to t\bar{t}$, $t\bar{t}\to
W^+W^-$, $t\bar{t}\to VV$ and $t\bar{t}\to ZH$ (where $H$ is the SM
Higgs field) at arbitrary center-of-mass energies.  The most stringent
bounds are obtained from $W^+W^-$ production in $t$--$\bar{t}$
annihilation.  We find
\begin{eqnarray}
\left\vert \Delta F^\gamma_{1V,A}(0)\right\vert & \leq & {96\pi\over\sqrt{6}
G_F}\,{1\over\sin^2\theta_W}\,{1\over \Lambda_{FF}^2}\approx \left({6.78~{\rm
TeV}\over \Lambda_{FF}}\right)^2,\label{eq:unit1}\\[2.mm]
\left\vert \Delta F^Z_{1V,A}(0)\right\vert & \leq & {96\pi\over\sqrt{6}
G_F}\,{1\over\sin\theta_W\cos\theta_W}\,{1\over \Lambda_{FF}^2}\approx
\left({5.01~{\rm TeV}\over \Lambda_{FF}}\right)^2,\label{eq:unit2}\\[2.mm]
\left\vert \Delta F^\gamma_{2V,A}(0)\right\vert & \leq & {128\sqrt{2}\pi\over
\sin^2\theta_W}\,{m_t\over G_F}\,{1\over \Lambda_{FF}^3}\approx
\left({3.35~{\rm TeV}\over \Lambda_{FF}}\right)^3,\label{eq:unit3} \\[2.mm]
\left\vert \Delta F^Z_{2V,A}(0)\right\vert & \leq & {128\sqrt{2}\pi\over
\sin\theta_W\cos\theta_W}\,{m_t\over G_F}\,{1\over \Lambda_{FF}^3}\approx
\left({2.75~{\rm TeV}\over \Lambda_{FF}}\right)^3,\label{eq:unit4}
\end{eqnarray}
where $G_F$ is the Fermi constant and $\theta_W$ is the weak mixing
angle.  We use a top quark mass of 178~GeV~\cite{topmass} in
Eqs.~(\ref{eq:unit1}--\ref{eq:unit4}).  Our results for $\Delta
F^Z_{1V,A}(0)$ are consistent with those obtained in
Ref.~\cite{Hosch:1996wu}.  For a step-function form factor, the bounds
on $\Delta F^Z_{1V,A}(0)$ ($\Delta F^Z_{2V,A}(0)$) in
Eqs.~(\ref{eq:unit1}) and~(\ref{eq:unit2}) (Eqs.~(\ref{eq:unit3})
and~(\ref{eq:unit4})) have to be divided by a factor~4 (16).

\subsection{Present Experimental Limits}
\label{sec:sec2c}

Although there are no current direct limits, precision measurements at
the $Z$ pole and the measured $b\to s\gamma$ branching ratio (BR)
provide indirect limits on the $ttV$ couplings.  Non-standard $ttZ$
couplings and the $tt\gamma$ dipole form factors, $F^\gamma_{2V,A}$,
contribute at one loop to the $\epsilon$ parameters of
Ref.~\cite{Altarelli:1991fk}.  The $b\to s\gamma$ BR gives additional
information on the $F^\gamma_{2V,A}$ couplings.  Non-standard $ttV$
coupling contributions to the $\epsilon$ parameters are divergent
unless the couplings' momentum dependence is properly taken into
account.  As discussed in Sec.~\ref{sec:sec2b}, one usually
regularizes the divergent integrals by assuming the form factors to be
of step-function form ($\theta(x)$ is the step-function):
\begin{equation}
\Delta F^V_{iV,A}(k^2)=\Delta F^V_{iV,A}(0)\,\theta(\Lambda^2-k^2).
\end{equation}
Extracting information on
anomalous couplings from loop observables assumes that no other
sources of new physics contribute to these observables.

Non-standard $ttZ$ vector and axial vector couplings, $\Delta
F^Z_{1V,A}$, are mostly constrained by the parameters $\epsilon_1$ and
$\epsilon_b$, which are closely related to the $\rho$ parameter and
the $Z\to b\bar{b}$ decay width. The terms proportional to $\Delta
F^Z_{1V,A}$ which contribute to $\epsilon_2$ and $\epsilon_3$ are
suppressed by a factor $m_W^2/m_t^2$ (where $m_W$ is the mass of the
$W$ boson) relative to those which appear in $\epsilon_1$ and
$\epsilon_b$. Using the expressions given in Ref.~\cite{Larios:1999au}
combined with the most recent experimental
results~\cite{Altarelli:2004mr} and SM
predictions~\cite{Altarelli:1997et} for the $\epsilon$ parameters, and
assuming that the couplings $\Delta F^Z_{1V,A}$ are real, we obtain
\begin{eqnarray}
-0.044 \leq &
-\Delta F^Z_{1A}(0) \left( 1 + 0.842\,\Delta F^Z_{1A}(0) \right) \,
\log \left( \frac{\Lambda^2}{m_t^2} \right) & \leq 0.065 \, , 
\label{eq:eps1}\\[2.mm]
-0.029 \leq &
-\left( \Delta F^Z_{1A}(0) - \frac{3}{5}\Delta F^Z_{1V}(0) \right)
\log \left( \frac{\Lambda^2}{m_t^2} \right) & \leq 0.143 \, .
\label{eq:eps2}
\end{eqnarray}
For $\Lambda={\cal O}(1$~TeV), Eqs.~(\ref{eq:eps1},\ref{eq:eps2})
constrain $\vert\Delta F^Z_{1V,A}(0)\vert$ to be less than a few
percent.

The effect of the magnetic dipole moment couplings $F^V_{2V}$ on the
$\epsilon$ parameters was analyzed in Ref~\cite{Eboli:1997kd}.  It
turns out that $F^V_{2V}$ affects only $\epsilon_2$ and $\epsilon_3$
and that these parameters constrain only a combination of
$F^\gamma_{2V}$ and $F^Z_{2V}$. From the most recent experimental
results and theoretical predictions for these parameters, one obtains:
\begin{eqnarray}
-0.92 \leq & -\left( F^\gamma_{2V}(0) + 1.83\, F^Z_{2V}(0) \right)
\log \left( \frac{\Lambda^2}{m_t^2} \right) & \leq 0 \, ,
\label{eq:eps3} \\[2.mm] 
-1.08 \leq & -\left( F^\gamma_{2V}(0) + 1.83\, F^Z_{2V}(0) \right)
\log \left( \frac{\Lambda^2}{m_t^2} \right) & \leq 1.92 \, ,
\label{eq:eps4}
\end{eqnarray}
where again we have assumed real $F^V_{2V}$.  If only one of the
couplings is allowed to deviate from its SM value,
Eqs.~(\ref{eq:eps3}) and~(\ref{eq:eps4}) yield $\left\vert
F^\gamma_{2V}(0)\right\vert\lesssim 0.3$ and $\left\vert
F^Z_{2V}(0)\right\vert\lesssim 0.2$ for $\Lambda=1$~TeV.  The effect
of the electric dipole moment couplings $F^V_{2A}$ on the $\epsilon$
parameters has not been studied so far.

Bounds on $F^\gamma_{2V,A}$ from $b\to s\gamma$ data can easily be
estimated from Refs.~\cite{Larios:1999au} and~\cite{Hewett:1993em}.
The latest CLEO and BELLE measurements of the $b\to s\gamma$ BR give
$BR(b\to s\gamma)=(3.3\pm 0.4)\cdot
10^{-4}$~\cite{Eidelman:2004wy}. The SM predicts $BR(b\to
s\gamma)=(3.4\pm 0.5\pm 0.4)\cdot 10^{-4}$~\cite{Neubert:2004dd},
where the first error is an estimate of the perturbative
uncertainties, and the second reflects uncertainties in the input
parameters. Adding the experimental and theoretical uncertainties in
quadrature, we find:
\begin{equation}
-0.39 \leq   1.94\,{\rm Re}(F^\gamma_{2V}(0))
           + 0.68\,{\rm Im}(F^\gamma_{2A}(0)) 
           + 0.45 \left\vert F^\gamma_{2V}(0)\right\vert^2
           + 0.056\left\vert F^\gamma_{2A}(0)\right\vert^2
\leq 1.11 \, .
\end{equation}
Assuming that $F^\gamma_{2V}$ and $F^\gamma_{2A}$ are real couplings,
and varying only one coupling at a time, one finds that $-0.2\leq
F^\gamma_{2V}(0)\leq 0.5$\footnote{A second solution, $-5.9\leq
F^\gamma_{2V}(0)\leq -4.1$ is clearly inconsistent with LEP data (see
Eqs.~(\ref{eq:eps3}) and~(\ref{eq:eps4})).} and
$|F^\gamma_{2A}(0)|\leq 4.5$.

The $tt\gamma$ vector and axial vector couplings are not constrained
by any current data.

\section{\boldmath{${t\bar{t}\gamma}$} Production}
\label{sec:sec3}

For {${t\bar{t}\gamma}$} production, as well as the $t\bar{t}Z$ process 
considered in the next section, we assume the Tevatron (LHC) to be 
operating at $\sqrt{s}=2.0\,(14)$~TeV.

\subsection{Signal}
\label{sec:sec3a}

The process $p\,p\hskip-7pt\hbox{$^{^{(\!-\!)}}$}\to t\bar{t}\gamma$
followed by $t\to Wb$ leads either to a
$\gamma\ell\nu_\ell\ell'\nu_{\ell'}b\bar{b}$ final state if both $W$
bosons decay leptonically, to a $\gamma\ell\nu_\ell b\bar{b}jj$ final
state if one $W$ decays leptonically and the other decays hadronically, 
or to a $\gamma b\bar{b}+4j$ final state if both $W$ bosons decay
hadronically.  The $\gamma b\bar{b}+4j$ final state has the largest
BR.  However, it is plagued by a large QCD background, so we ignore
it.  The dilepton final state, although less contaminated by
background, has a BR about a factor~6 smaller than that of the
so-called lepton+jets mode.  In the following, we therefore
concentrate on this last process:
\begin{equation}
p\,p\hskip-7pt\hbox{$^{^{(\!-\!)}}$} \to \gamma\ell\nu_\ell b\bar{b}jj
\end{equation}
with $\ell=e\,,\mu$ ($\tau$ leptons are ignored).  We assume that both
$b$ quarks are tagged with a combined efficiency of
$\epsilon_b^2=25\%$ ($\epsilon_b^2=40\%$) at the Tevatron (LHC),
unless explicitly stated otherwise.

We perform our calculation for general $tt\gamma$ couplings of the
form of Eq.~(\ref{eq:anomvertex}).  As we shall see, at both the
Tevatron and the LHC, photon transverse momenta of at most a few
hundred GeV are accessible.  The scale of new physics responsible for
anomalous $tt\gamma$ couplings is expected to be of ${\cal O}(1$~TeV)
or higher.  Form factor effects will thus be small and are therefore
neglected in the following.  We also assume that all $tt\gamma$
couplings are real.  We otherwise assume the SM to be valid.  In
particular, we assume that the $bb\gamma$ coupling is that of the 
SM.  Our analysis for $F^\gamma_{1V}$ thus differs from that of
Ref.~\cite{Baur:2001si} for the top quark electric charge ($Q_t$)
measurement.  That study assumed that $Q_t$ is related to the
$b$ quark charge, $Q_b$, and $W$ boson charge, $Q_W=\pm 1$, by
$Q_t=Q_b+Q_W$.

Our calculation includes top quark and $W$ decays with full spin
correlations and finite width effects.  All Feynman diagrams
contributing to the lepton+jets final state are included, i.e. besides
$t\bar{t}\gamma$ production, we automatically take into account 
top quark pair production where one of the top quarks decays 
radiatively, $t\to Wb\gamma$.  Subsequently, we will refer to this 
process simply as ``$t\bar{t}\gamma$ production'' and it is implied 
that it automatically includes any contribution from $t\bar{t}$ 
production where one of the top quarks undergoes radiative decay.  
To ensure gauge invariance of the SM cross section, we use the 
so-called overall-factor scheme of Ref.~\cite{Baur:1991pp}, as 
implemented for $t\bar{t}V$ production in Ref.~\cite{Maltoni:2002jr}.

All signal and background cross sections in this paper are computed
using CTEQ6L1~\cite{Pumplin:2002vw} parton distribution functions with
the strong coupling constant evaluated at leading order and
$\alpha_s(m_Z^2)=0.130$, where $m_Z$ is the $Z$-boson mass.  The top
quark mass is assumed to be $m_t=178$~GeV~\cite{topmass}.  All signal
cross sections in this paper are calculated for factorization and
renormalization scales equal to $m_t$.

The acceptance cuts for $\gamma\ell\nu_\ell b\bar{b}jj$ events at the
Tevatron (LHC) are
\begin{eqnarray}
\label{eq:cuts1}
\nonumber &
\sla{p}_T>20~{\rm GeV} \; , \\
\nonumber &
p_T(b) > 15~(20)~{\rm GeV} \; , \qquad
|\eta(b)| < 2.0~(2.5) \; , \qquad
\Delta R(b,b) > 0.4 \; , \\
 &
p_T(j) > 20~{\rm GeV} \; , \qquad
|\eta(j)| < 2.5 \; , \qquad
\Delta R(j,j) > 0.4 \; , \qquad
\Delta R(j,b) > 0.4 \; , \\
\nonumber &
p_T(\gamma) > 10~(30)~{\rm GeV} \; , \qquad
|\eta(\gamma)| < 2.5 \; , \qquad
\Delta R(\gamma,j) > 0.4 \; , \qquad
\Delta R(\gamma,b) > 1.0 \; , \\
\nonumber &
p_T(\ell) > 15~{\rm GeV} \; , \quad
|\eta(\ell)| < 2.5 \; , \quad
\Delta R(\ell,\gamma) > 0.4 \; ,\quad
\Delta R(\ell,j) > 0.4 \; , \quad
\Delta R(\ell,b) > 0.4 \; ,
\end{eqnarray}
where $\Delta R=[(\Delta\Phi)^2+(\Delta\eta)^2]^{1/2}$ is the
separation in pseudorapidity -- azimuth space and $\sla{p}_T$ is the
missing transverse momentum originating from the neutrino which
escapes undetected.  We include minimal detector effects via Gaussian
smearing of parton momenta according to CDF~\cite{cdfii} and
CMS~\cite{cms} expectations, and take into account the $b$ jet energy
loss via a parameterized function.

Since we are interested in photon emission from top quarks, we would
like to suppress radiation from $W$ decay products, as well as
emission from $b$ quarks and from initial-state quarks.  The large
$\Delta R(\gamma,b)$ cut in Eq.~(\ref{eq:cuts1}) reduces photon
radiation from the $b$ quarks.  Photon emission from $W$ decay
products can essentially be eliminated by requiring that
\begin{equation}\label{eq:cuts2}
m(jj\gamma) > 90~{\rm GeV} \qquad {\rm and} \qquad
m_T(\ell\gamma;\sla{p}_T) > 90~{\rm GeV,} 
\end{equation}
where $m(jj\gamma)$ is the invariant mass of the $jj\gamma$ system.
The variable $m_T(\ell\gamma;\sla{p}_T)$ is the $\ell\gamma\sla{p}_T$
cluster transverse mass, given by
\begin{equation}
m_T^2(\ell\gamma;\sla{p}_T)=\left(\sqrt{p_T^2(\ell\gamma)+m^2(\ell\gamma)}
+ \sla{p}_T\right)^2-\left(\vec{p}_T(\ell\gamma)+
\vec{\sla{p}}_T\right)^2\, ,
\end{equation}
where $p_T(\ell\gamma)$ and $m(\ell\gamma)$ are the transverse
momentum and invariant mass of the $\ell\gamma$ system, respectively.
The $\ell\gamma\sla{p}_T$ cluster transverse mass peaks sharply at
$m_W$.  It is difficult to suppress radiation from the initial state
quarks without simultaneously reducing the signal cross section by an
equal amount.  Fortunately this is not a problem at the LHC, where
gluon fusion dominates.

In addition to the cuts listed in Eqs.~(\ref{eq:cuts1})
and~(\ref{eq:cuts2}), we require that the event is consistent either
with $t\bar{t}\gamma$ production, or with $t\bar{t}$ production with
radiative top decay.  This will reduce the singly-resonant and 
non-resonant backgrounds, and is accomplished by selecting events 
which satisfy either
\begin{equation}\label{eq:cuts5}
m_T(b_{1,2}\ell;\sla{p}_T) < m_t + 20~{\rm GeV}
\qquad {\rm and} \qquad
m_t - 20~{\rm GeV} < m(b_{2,1}jj) < m_t + 20~{\rm GeV} ,
\end{equation}
\begin{equation}
m_T(b_{1,2}\ell\gamma;\sla{p}_T) < m_t + 20~{\rm GeV}
\qquad {\rm and} \qquad 
m_t - 20~{\rm GeV} < m(b_{2,1}jj) < m_t + 20~{\rm GeV} ,
\end{equation}
or
\begin{equation}\label{eq:cuts3}
m_T(b_{1,2}\ell;\sla{p}_T) < m_t + 20~{\rm GeV}
\qquad {\rm and} \qquad
m_t - 20~{\rm GeV} < m(b_{2,1}jj\gamma) < m_t + 20~{\rm GeV} .
\end{equation}
Here, $b_1,b_2=b,\bar{b}$, and $b_1\neq b_2\,$.

Imposing the cuts listed in Eqs.~(\ref{eq:cuts1}--\ref{eq:cuts3}), and
before taking into account particle identification efficiencies, we
obtain a cross section of about 5~fb (82~fb) at the Tevatron
(LHC).  The total integrated luminosity one hopes to achieve at the
Tevatron in Run~II is between 4~and 8~fb$^{-1}$.  While this will not
be sufficient for a precision measurement of the $tt\gamma$ couplings,
it may offer a chance for a first test of these couplings.  At the LHC,
with 300~fb$^{-1}$, one expects several thousand signal events which
should make it possible to precisely determine the $tt\gamma$
couplings if the background can be controlled.

\subsection{Background Processes}
\label{sec:sec3b}

The most important irreducible background processes that remain after 
imposing the cuts described in Sec.~\ref{sec:sec3a}, are 
$t(\to b\ell^+\nu)\bar{b}\gamma jj$, $\bar{t}(\to\bar{b}\ell^-\bar\nu) 
b\gamma jj$, $t(\to bjj)\bar{b}\ell^-\bar\nu\gamma$ and
$\bar{t}(\to\bar{b}jj)b\ell^+\nu\gamma$ production, and the
non-resonant process $p\,p\hskip-7pt\hbox{$^{^{(\!-\!)}}$} \to
W(\to\ell\nu)\gamma b\bar{b}jj$.  The single-top processes will be
collectively denoted as ``$(t\bar{b}\gamma + \bar{t}b\gamma)+X$
production'' in the following.  We calculate the irreducible
background processes at leading order in QCD including the full set of
contributing Feynman diagrams using {\tt
MADEVENT}~\cite{Maltoni:2002qb}.  $W(\to\ell\nu\gamma)b\bar{b}jj$
production, as well as $t\bar{b}jj$, $\bar{t}bjj$,
$t\bar{b}\ell^-\bar\nu$ and $\bar{t}b\ell^+\nu$ production where the
top quark decays radiatively, are strongly suppressed by the cuts of
Eqs.~(\ref{eq:cuts2}--\ref{eq:cuts3}) and therefore not considered.

There are also several reducible backgrounds resulting from light jets
faking either $b$ jets or photons, or from $Z$ bosons where one of the
leptons in $Z\to\ell^+\ell^-$ is lost and fakes missing transverse
momentum.  To estimate these backgrounds we assume the probability
of a light jet to be misidentified as a $b$ jet
to be~\cite{flera,Gianotti:2002xx}
\begin{equation}\label{eq:bmisid}
P_{j\to b} = 1/100~(1/140)
\end{equation}
at~the~Tevatron~(LHC).  For the probability of a jet to fake a photon,
$P_{j\to\gamma}$, at the Tevatron we use the result obtained by CDF
for $10~{\rm GeV}\leq p_T(\gamma)\leq 25$~GeV in the measurement of
the $W\gamma$ and $Z\gamma$ cross sections~\cite{cdfprob}, and
conservatively assume that $P_{j\to\gamma}$ is constant for
$p_T(\gamma)\geq 25$~GeV:
\begin{equation}\label{eq:pjg}
P_{j\to\gamma}=
\begin{cases}
a\cdot e^{-b\cdot p_T(\gamma)} & 
\text{for $10~{\rm GeV}\leq p_T(\gamma)\leq 25$~GeV,} \\
7\cdot 10^{-4} & 
\text{for $p_T(\gamma)\geq 25$~GeV,}
\end{cases}
\end{equation}
with $a=0.0079$ and $b=0.097~{\rm GeV}^{-1}$.  The D\O\ Collaboration
obtained a similar result~\cite{d0prob}.  Expectations for the
probability to misidentify a light jet as a photon at the LHC vary
between $P^{lo}_{j\to\gamma}=1/2500$ and
$P^{hi}_{j\to\gamma}=1/1600$~\cite{cms,atlas_tdr,schwem,Abdullin:1998er}. In
the following we take the conservative route and  use the more
pessimistic estimate 
$P_{j\to\gamma}=1/1600$ for all numerical studies at LHC.

The potentially most dangerous reducible background is $t\bar{t}j$
production where one of the jets in the final state fakes a photon.
We calculate this using exact $W^+W^-b\bar{b}j$ matrix elements,
including spin correlations for the $W$ decays.  However, gluon
radiation from the $W$ decay products is not included.  For the cuts
used here, Eqs.~(\ref{eq:cuts1}--\ref{eq:cuts3}), this should be an
excellent approximation to the full process
$p\,p\hskip-7pt\hbox{$^{^{(\!-\!)}}$} \to \ell\nu b\bar{b}+3$~jets.

\begin{figure}[t!]
\begin{center}
\includegraphics[width=12.5cm]{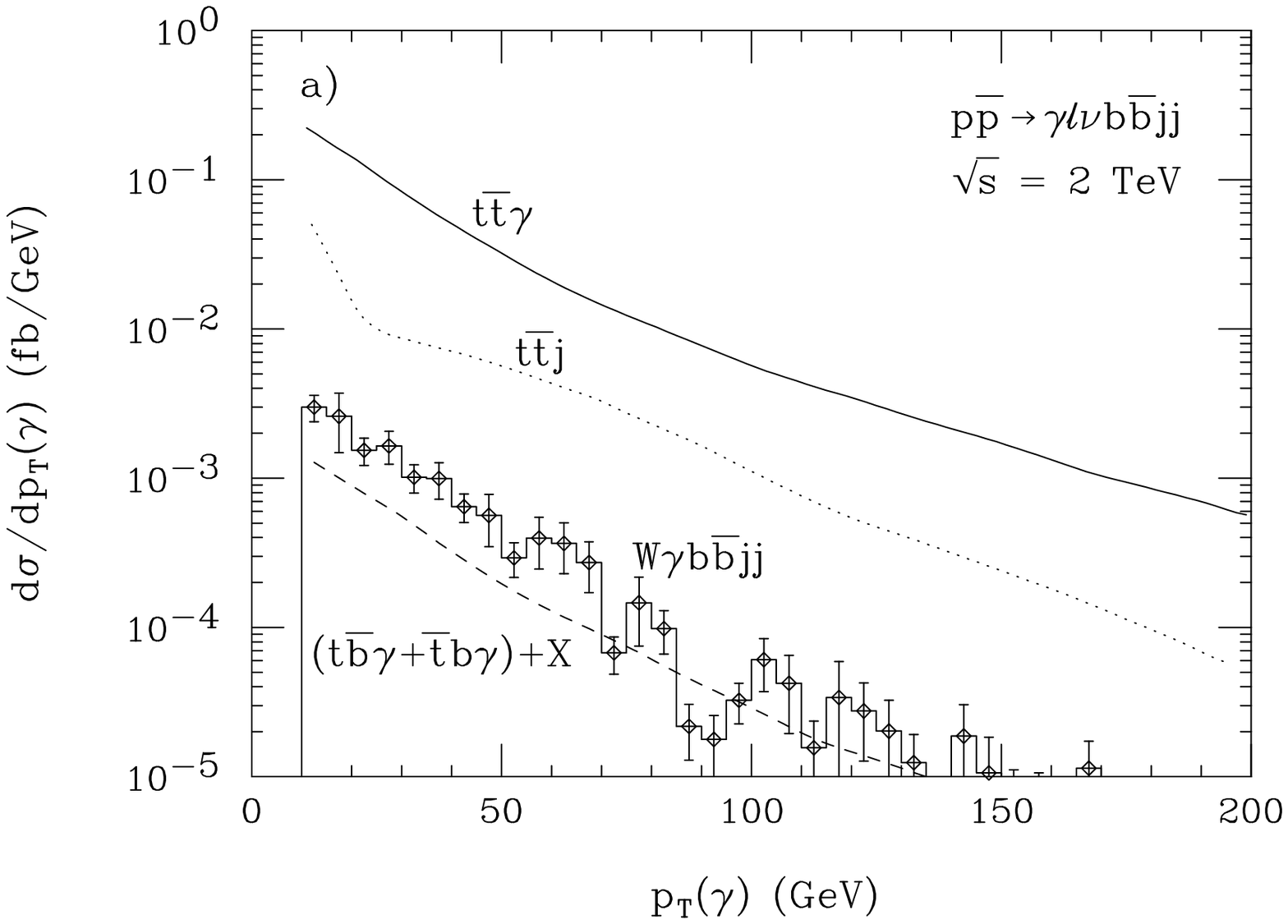} \\
\includegraphics[width=12.5cm]{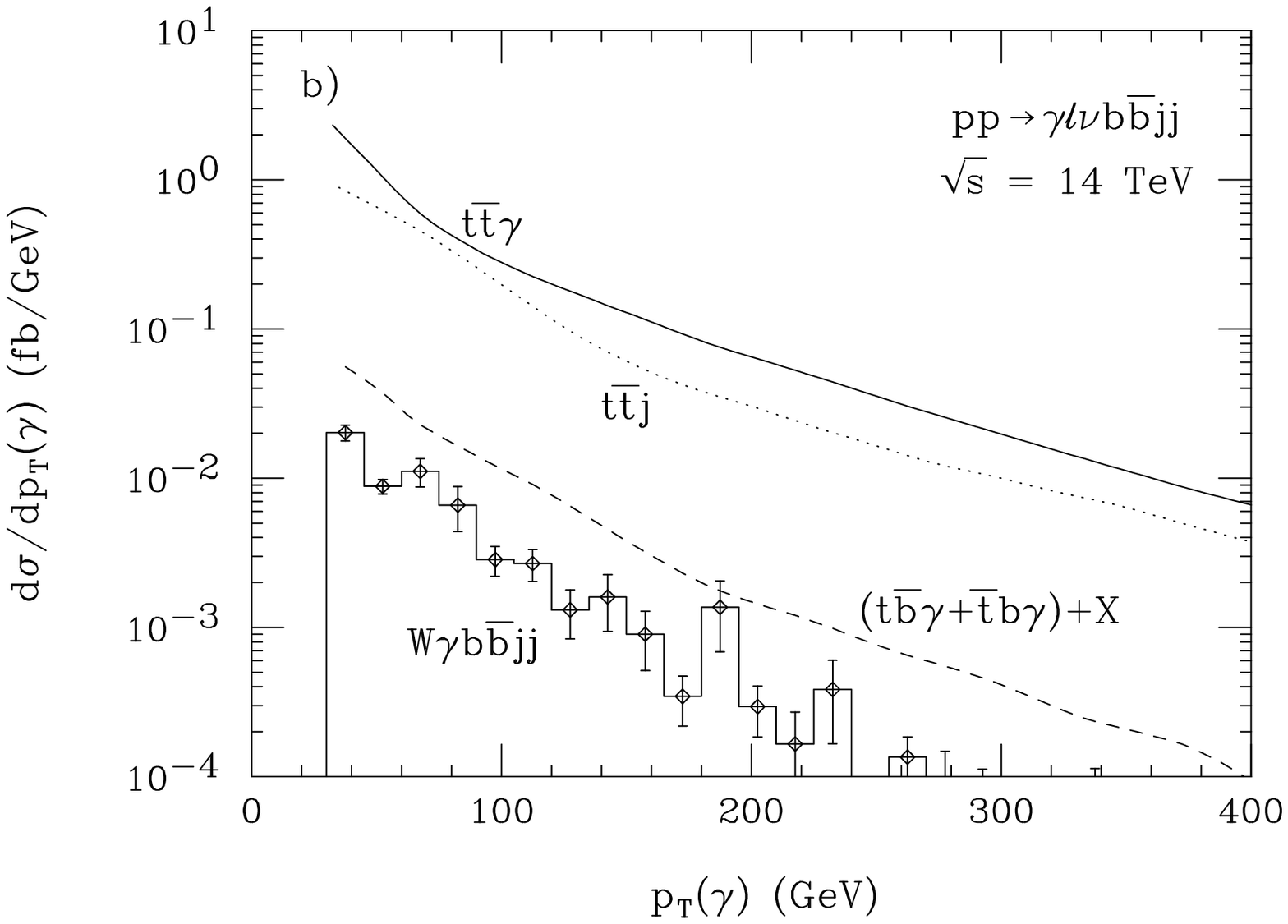} 
\vspace*{2mm}
\caption[]{The differential cross sections as a function of the photon 
transverse momentum for $\gamma\ell\nu_\ell b\bar{b}jj$ production at
(a) Tevatron Run II and (b) LHC.  Shown are the SM predictions for
$t\bar{t}\gamma$ production (including radiative top decays in
$t\bar{t}$ events, solid line), the $t\bar{t}j$ background where one
jet is misidentified as a photon (dotted line), the background from
single-top production processes (dashed line), and the $W\gamma
b\bar{b}jj$ background (histogram). The cuts imposed are listed in
Eqs.~(\ref{eq:cuts1}--\ref{eq:cuts3}). The photon misidentification
probabilities used are described in the text.  No particle ID
efficiencies are included here.}
\label{fig:fig1}
\vspace{-7mm}
\end{center}
\end{figure}
In Fig.~\ref{fig:fig1} we show the photon transverse momentum
distributions of the $t\bar{t}\gamma$ signal (solid curve), the
$t\bar{t}j$ background (dotted line), the background from single top
production processes (dashed line), and the $W\gamma b\bar{b}jj$
background (histogram). There are several thousand Feynman diagrams
contributing to $W\gamma b\bar{b}jj$ production.  Numerical evaluation
of these helicity amplitudes is very time consuming. We therefore show
the $W\gamma b\bar{b}jj$ differential cross section in form of a
histogram, where the error bars represent the statistical uncertainty
of the Monte Carlo integration. The $t\bar{t}j$ background is seen to
be a factor~2 to~10 smaller than the $t\bar{t}\gamma$ signal for the
jet -- photon misidentification probabilities used. The sharp kink in
the $t\bar{t}j$ differential cross section at the Tevatron is due to
the functional form of $P_{j\to\gamma}$ (see Eq.~(\ref{eq:pjg})). The
$(t\bar{b}\gamma +
\bar{t}b\gamma)+X$ and $W\gamma b\bar{b}jj$ backgrounds both are found
to be more than an order of magnitude smaller the $t\bar{t}j$
background.

The numerical results shown in Fig.~\ref{fig:fig1} and all subsequent
figures which display differential cross sections represent cross
sections after selection cuts but {\sl before} any particle 
identification efficiencies are taken into account.

It should be noted that the cross sections of the $t\bar{t}j$, the
$(t\bar{b}\gamma + \bar{t}b\gamma)+X$, and the $W\gamma b\bar{b}jj$
backgrounds depend significantly on the choice of factorization and
renormalization scales, $\mu_F$ and $\mu_R$, which were taken to be
$\mu_F=\mu_R=m_t$.  Including next-to-leading oder (NLO) corrections
in most cases significantly reduces the scale dependence of a process.
Unfortunately the NLO QCD corrections are not presently known for any 
of the background processes.  However, at least the $t\bar{t}j$ rate
should eventually be well-measured in data.

Other reducible background sources are $t\bar{b}+3$~jet,
$\bar{t}b+3$~jet, $t\bar{b}\ell^-\bar\nu j$, $\bar{t}b\ell^+\nu j$ and
$Wb\bar{b}+3$~jet production, where one jet fakes a photon; $W\gamma
+4$~jet production where where two jets are misidentified as $b$ jets;
$WW+3$~jet production where one jet fakes a photon and two jets are
misidentified as $b$ jets; and $Z\gamma b\bar{b}jj$ production where
one of the leptons from the $Z$ decay is lost (we implicitly mean 
$Z/\gamma^*$ whenever the final state is dileptons).  We find the 
combined cross section for the single-top + jet(s) processes to be 
about a factor~10 smaller than that for $(t\bar{b}\gamma + 
\bar{t}b\gamma)+X$ production; similarly for the $Wb\bar{b}+3$~jet 
background.  Since $P_{j\to b}$ is very small {\it and} we require 
two tagged $b$ jets, the background from $W\gamma +4$~jet production 
where two jets are misidentified as $b$ jets is negligible; so is the 
$WW+3$~jet background.  It should be noted that for a 
luminosity-upgraded LHC (SLHC), $P_{j\to b}$ may dramatically increase 
with as many as one in four light jets being misidentified as a $b$
quark~\cite{Gianotti:2002xx}.  In this case, the $W\gamma +4$~jet
cross section may be of the same order as that of $W\gamma
b\bar{b}jj$.

Reducible $Z\gamma b\bar{b}jj$ production contributes to the
background if one of the leptons from $Z/\gamma^*$ decay is missed.
We consider a lepton to be missed if it has $p_T<10$~GeV or
$|\eta|>2.5$.  If the lepton is within a cone of $\Delta R<0.2$ from a
detected lepton and has $1~{\rm GeV}<p_T<10$~GeV, the detected lepton
is not considered isolated and we reject the event.  In order to avoid
the collinear singularity when the missed lepton is collinear with an
observed lepton (which is relevant only if the the missed lepton has
$p_T<1$~GeV), we retain finite lepton masses in the calculation.

With several$\times10^4$ Feynman diagrams contributing,
$p\,p\hskip-7pt\hbox{$^{^{(\!-\!)}}$} \to Z\gamma b\bar{b}jj$ is
sufficiently complicated that it requires approximation.  To estimate
the $Z\gamma b\bar{b}jj$ background we use a procedure similar to that
described in Ref.~\cite{Maltoni:2002jr}.  We first calculate the ratio
of the $W\gamma b\bar{b}$ and $W\gamma b\bar{b}jj$ cross sections.  We
then calculate the $Z\gamma b\bar{b}$ cross section (including
$\gamma^*\to\ell^+\ell^-$ interference) where one of the leptons from
$Z/\gamma^*$ is missed, and scale it by the $W\gamma b\bar{b}jj$ and
$W\gamma b\bar{b}$ cross section ratio.  Since they entail QCD
radiation from very similar subprocesses, the $Z\gamma
b\bar{b}jj/Z\gamma b\bar{b}$ and $W\gamma b\bar{b}jj/W\gamma b\bar{b}$
cross section ratios are expected to be approximately equal.  At the
Tevatron (LHC), we find that the estimated $Z\gamma b\bar{b}jj$ cross
section is about a factor~7 (2) smaller than that of $W\gamma
b\bar{b}jj$.

In addition to the backgrounds considered so far, $\gamma\ell\nu_\ell
b\bar{b}jj$ events (or their fakes) may also be produced in double
parton scattering (DPS), or from multiple interactions occurring from
separate $p\,p\hskip-7pt\hbox{$^{^{(\!-\!)}}$}$ collisions in the same
bunch crossing at high-luminosity running.  In principle, one can
identify multiple interactions by a total visible energy measurement
or by tracing some final particle tracks back to distinctly separate
primary vertices, but this may not always be possible in practice.  To
estimate the cross sections from DPS and multiple interactions, we use
the approximation outlined in Ref.~\cite{BCHP}.  At the LHC, the cross
section from overlapping events is about a factor of two larger than
that from DPS.  At the Tevatron, for a luminosity of ${\cal L} = 
10^{32}~{\rm cm^{-2}\,s^{-1}}$, DPS dominates.  The resulting background 
arises predominantly from the overlap of a $t\bar{t}$ event and a 
two-jet event, wherein one jet is misidentified as a photon and the 
other is missed.  We estimate the cross section for this process to be
approximately 0.7~fb (0.01~fb) at the LHC (Tevatron), which is of the
same order or smaller than for $W\gamma b\bar{b}jj$.  The cross
sections for the SM signal and the most important background processes
are summarized in Table~\ref{tab:tab0}.
\begin{table}
\caption{Expected cross sections (fb) for the $\gamma\ell\nu_\ell 
b\bar{b}jj$ signal and the most important background processes at the 
Tevatron and the LHC for the cuts described in Sec.~\ref{sec:sec3a}.  
The photon misidentification probabilities used are described in the 
text.  No particle ID efficiencies are included.}
\label{tab:tab0}
\vspace{2mm}
\begin{tabular}{ccc}
process & Tevatron & LHC\\
\tableline 
signal & 4.9 & 81.7 \\
$t\bar{t}j$ & 0.78 & 45.7  \\
$(t\bar{b}\gamma+\bar{t}b\gamma)+X$ & 0.03 & 2.64 \\
$W\gamma  b\bar{b}jj$ & 0.07 & 0.89 \\
$Z\gamma b\bar{b}jj$ & 0.01 & 0.43 \\
$t\bar{t}\oplus jj$ & 0.01 & 0.7
\end{tabular}
\end{table}

As stated before, we require that both $b$ quarks be tagged. Requiring
only one tagged $b$ quark would result in a signal cross section
increase of a factor $(2/\epsilon_b-1)$.  This larger signal rate
comes at the expense of an increased background and a reduced
acceptance.  In events where one of the $b$ quarks is not tagged,
photon radiation off the untagged $b$ quark cannot be suppressed by a
larger $\Delta R$ cut.  Furthermore, to suppress the contributions
from radiative $W$ decay, the invariant mass cut on the $jj\gamma$
system in Eq.~(\ref{eq:cuts2}) has to be imposed on all three possible
$jj\gamma$ combinations. This reduces the signal cross section by
almost a factor~2.  In addition, for events with only one $b$ tag, the
background will be larger. The $t\bar{t}j$ background increases by
roughly $30\%$ relative to the signal.  The
$(t\bar{b}\gamma+\bar{t}b\gamma)+X$ and $W\gamma b\bar{b}jj$
backgrounds increase due to the larger combinatorial background from
grouping jets, the tagged $b$ quark and the $\ell\nu$ system into
$b\ell\nu(\gamma)$, $jjj(\gamma)$, $j\ell\nu(\gamma)$ and
$bjj(\gamma)$ systems which are compatible with (radiative) top decay.
Detailed calculations are needed for a quantitative estimate of the
increase of these backgrounds.  Finally, the $W\gamma+4$~jet and
$WW+3$~jet backgrounds increase by about two orders of magnitude due
to the much higher probability that only one (instead of two) light
jet is mistagged as a $b$ quark.  Nevertheless, they are still
expected to be far smaller than the $W\gamma b\bar{b}jj$
background. Since the single-$b$-tagged final state is less ``clean''
than that where both $b$ quarks are identified, we do not consider it
in detail here.

\subsection{Signatures for anomalous $tt\gamma$ couplings}
\label{sec:sec3c}

\begin{figure}[t!]
\begin{center}
\includegraphics[width=13cm]{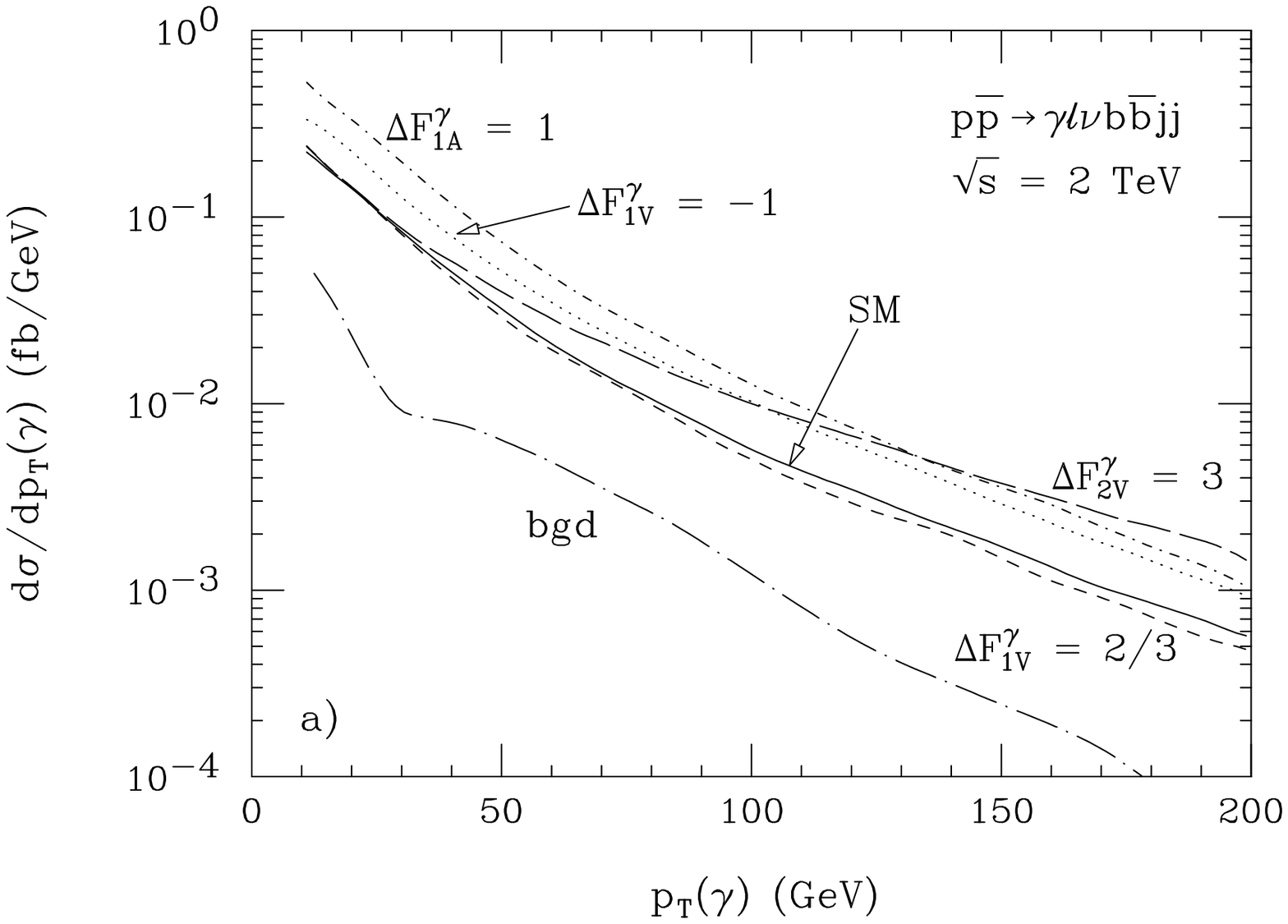} \\
\includegraphics[width=13cm]{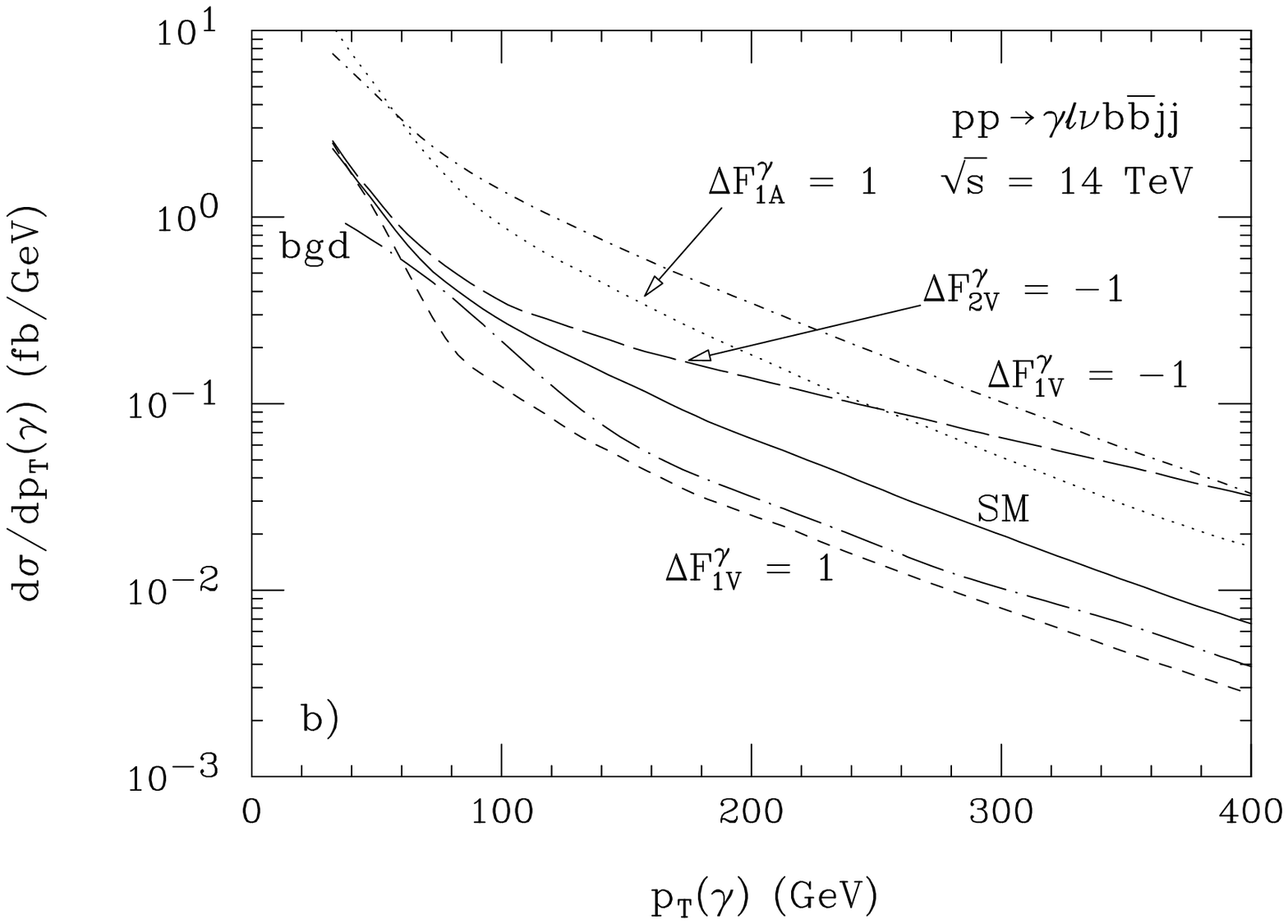} 
\vspace*{2mm}
\caption[]{The differential cross sections as a function of the photon 
transverse momentum for $\gamma\ell\nu_\ell b\bar{b}jj$ production at
(a) Tevatron Run II and (b) LHC.  Shown are the SM predictions for
$t\bar{t}\gamma$ production (including radiative top decays in
$t\bar{t}$ events, solid line), the combined $t\bar{t}j$, $W\gamma
b\bar{b}jj$ and $(t\bar{b}\gamma+\bar{t}b\gamma)+X$ background
(long-dashed-dotted line), and the predictions for several
non-standard $tt\gamma$ couplings. Only one coupling at a time is
allowed to deviate from its SM value. The cuts imposed are listed
in Eqs.~(\ref{eq:cuts1}--\ref{eq:cuts3}). No particle ID
efficiencies are included here.}
\label{fig:fig2}
\vspace{-7mm}
\end{center}
\end{figure}

The photon transverse momentum distributions for
$p\,p\hskip-7pt\hbox{$^{^{(\!-\!)}}$} \to \gamma\ell\nu_\ell
b\bar{b}jj$ in the SM and for various anomalous $tt\gamma$ couplings,
together with the combined $p_T(\gamma)$ distribution of the
$t\bar{t}j$, $W\gamma b\bar{b}jj$ and the
$(t\bar{b}\gamma+\bar{t}b\gamma)+X$ backgrounds, are shown in
Fig.~\ref{fig:fig2}.
Only one coupling at a time is allowed to deviate from its SM
prediction. At the Tevatron, the $\gamma\ell\nu_\ell b\bar{b}jj$ cross
section is completely dominated by $q\bar{q}$ annihilation. As a
result, photon radiation off the initial state quarks constitutes an
irreducible background which limits the sensitivity of the photon
differential cross section to anomalous $tt\gamma$ couplings. This is
particularly pronounced for $F^\gamma_{1V}$. Even when the photon does
not couple to the top quark at all ($\Delta F^\gamma_{1V}=2/3$ with
all other $tt\gamma$ couplings vanishing; dashed line in
Fig.~\ref{fig:fig2}a), the cross section hardly differs from the SM
result. In contrast, at the LHC more than $75\%$ of the
$\gamma\ell\nu_\ell b\bar{b}jj$ cross section originates from gluon
fusion.  This results in a greatly-increased sensitivity of the
$p_T(\gamma)$ distribution to non-standard $tt\gamma$ couplings, which
is evident from Fig.~\ref{fig:fig2}b.

Non-standard vector and axial vector couplings yield a transverse
momentum distribution for the photon with high-$p_T$ behavior similar
to that in the SM.  At low photon transverse momenta, however, the
shape of the $p_T$ distribution for SM and anomalous couplings
differs.  This is most easily noticed for $\Delta F^\gamma_{1V}=1$ in
Fig.~\ref{fig:fig2}b.  The change in shape at low $p_T$ is due to
radiative top decays which can contribute only in this region.
Non-standard and SM helicity amplitudes interfere differently for
$t\bar{t}\gamma$ production and $t\bar{t}$ events where one of the top
quarks decays radiatively, resulting in a shape change.  Since the
interference effects can be constructive or destructive, non-standard
vector or axial vector couplings can either increase or decrease the
signal cross section.

Terms in the helicity amplitudes proportional to the dipole form
factors $F^\gamma_{2V,A}$ grow like $m(t\gamma)/m_t$ at high energies.
Here, $m(t\gamma)$ is the invariant mass of the photon and the top
quark to which it couples.  This results in a transverse momentum
distribution of the photon which is considerably harder than that of a
non-standard vector or axial vector coupling.  The long-dashed curves
in Fig.~\ref{fig:fig2} show the photon $p_T$ distribution for
$F^\gamma_{2V}=3$ ($F^\gamma_{2V}=-1$) at the Tevatron (LHC).  For
equal coupling strengths, the numerical results obtained for
$F^\gamma_{2A}$ are almost identical to those found for
$F^\gamma_{2V}$.  To discriminate $F^\gamma_{2V}$ from
$F^\gamma_{2A}$, one can take advantage of the $CP$-violating nature
of $F^\gamma_{2A}$ and use the asymmetry $A^\mu_{cut}(p_{Tcut})$
introduced in Ref.~\cite{Grzadkowski:1997yi}.

Anomalous $tt\gamma$ couplings also affect the single resonant
$(t\bar{b}\gamma+\bar{t}b\gamma)+X$ background.  However, since the
$(t\bar{b}\gamma+\bar{t}b\gamma)+X$ background is small, this has
almost no effect on the overall signal to background ratio, so we do
not include the anomalous couplings in these backgrounds.

\section{\boldmath{${t\bar{t}Z}$} Production}
\label{sec:sec4}

The process $p\,p\hskip-7pt\hbox{$^{^{(\!-\!)}}$} \to t\bar{t}Z$ leads
to either ${\ell'}^+{\ell'}^-\ell\nu b\bar{b}jj$ or
${\ell'}^+{\ell'}^- b\bar{b}+4j$ final states if the $Z$-boson
decays leptonically and one of the $W$ bosons decays hadronically.
For both final states the leptonic $Z$ decay provides an efficient
trigger.  If the $Z$ boson decays into neutrinos and both $W$ bosons
decay hadronically, the final state consists of
$\sla{p}_Tb\bar{b}+4j$.  In this case one has to trigger on the
multijet system, similar to many supersymmetry searches.  For
$Z\to\bar\nu\nu$ and one of the $W$ bosons decaying leptonically, the
$t\bar{t}$ background swamps the signal.  Finally, for $Z\to jj
(b\bar{b})$, $t\bar{t}jj$ ($t\bar{t}b\bar{b}$) production constitutes
an overwhelming irreducible background.

In the following, we concentrate on the ${\ell'}^+{\ell'}^-\ell\nu
b\bar{b}jj$ and ${\ell'}^+{\ell'}^- b\bar{b}+4j$ final states,
which we henceforth refer to as the trilepton and dilepton channels
for brevity.  The ${\ell'}^+{\ell'}^-\ell\nu{\ell''}{\nu''}b\bar{b}$
channel, while experimentally cleaner, has a much smaller BR, so we
ignore it.  Due to the larger $Z\to\bar\nu\nu$ BR, the
$\sla{p}_Tb\bar{b}+4j$ channel cross section before cuts is about a
factor~3 larger than that for the trilepton and dilepton final states.
However, $t\bar{t}$ production with all-hadronic decays where one or
more jets are badly mismeasured, and $t\bar{t}W$ production where the
lepton from $W$ decay is lost, constitute potentially large
backgrounds.  For this reason, we also do not consider the
$\sla{p}_Tb\bar{b}+4j$ final state here.

The signal cross section calculation proceeds similar to that in
Sec.~\ref{sec:sec3}.  As in that case, form factor effects turn out to
unimportant and are ignored.  We assume real $ttZ$ couplings.  As with
$t\bar{t}\gamma$ we include all decay spin correlations and finite
width effects.  Here we also include off-shell photon interference
effects with $Z\to{\ell'}^+{\ell'}^-$.  We take into account all Feynman
diagrams contributing to the trilepton and dilepton final states,
including those 
where the final state $W$ boson couples to $\ell'$.  To ensure gauge
invariance of the SM result, we again use the overall-factor scheme.

\subsection{The {\boldmath $t\bar{t}Z$} trilepton final state}
\label{sec:sec4a}

In order to identify leptons, $b$ quarks, light jets and the missing
transverse momentum in ${\ell'}^+{\ell'}^-\ell\nu b\bar{b}jj$ events,
we impose the cuts listed in Eq.~(\ref{eq:cuts1}).  In addition, we
require that there is a same-flavor, opposite-sign lepton pair with
invariant mass near the $Z$ resonance,
\begin{equation}\label{eq:cuts4}
m_Z - 10~{\rm GeV} < m(\ell\ell) < m_Z + 10~{\rm GeV} .
\end{equation}
As a result of this final state signature requirement, $t\bar{t}Z$
production as observed is very insensitive to anomalous $tt\gamma$
couplings.  Since there is essentially no phase space for $t\to WZb$
decays ($BR(t\to WZb)\approx 3\cdot
10^{-6}$~\cite{Mahlon:1994us,Altarelli:2000nt}), this trilepton final
state arises only from $t\bar{t}Z$ production.  Thus, in addition to
the cuts listed in Eqs.~(\ref{eq:cuts1}) and~(\ref{eq:cuts4}), we
require that events satisfy Eq.~(\ref{eq:cuts5}), i.e. that the
$b\ell\nu$ and $bjj$ systems are consistent with top decay.

The main backgrounds contributing to the trilepton final state are
singly-resonant $(t\bar{b}Z+\bar{t}bZ)+X$ ($t\bar{b}Zjj$,
$\bar{t}bZjj$, $t\bar{b}Z\ell\nu$ and $\bar{t}bZ\ell\nu$) and
non-resonant $WZb\bar{b}jj$ production. As in the $t\bar{t}\gamma$
case, backgrounds from DPS and overlapping events are found to be
negligible.

\begin{figure}[t!]
\begin{center}
\includegraphics[width=15.5cm]{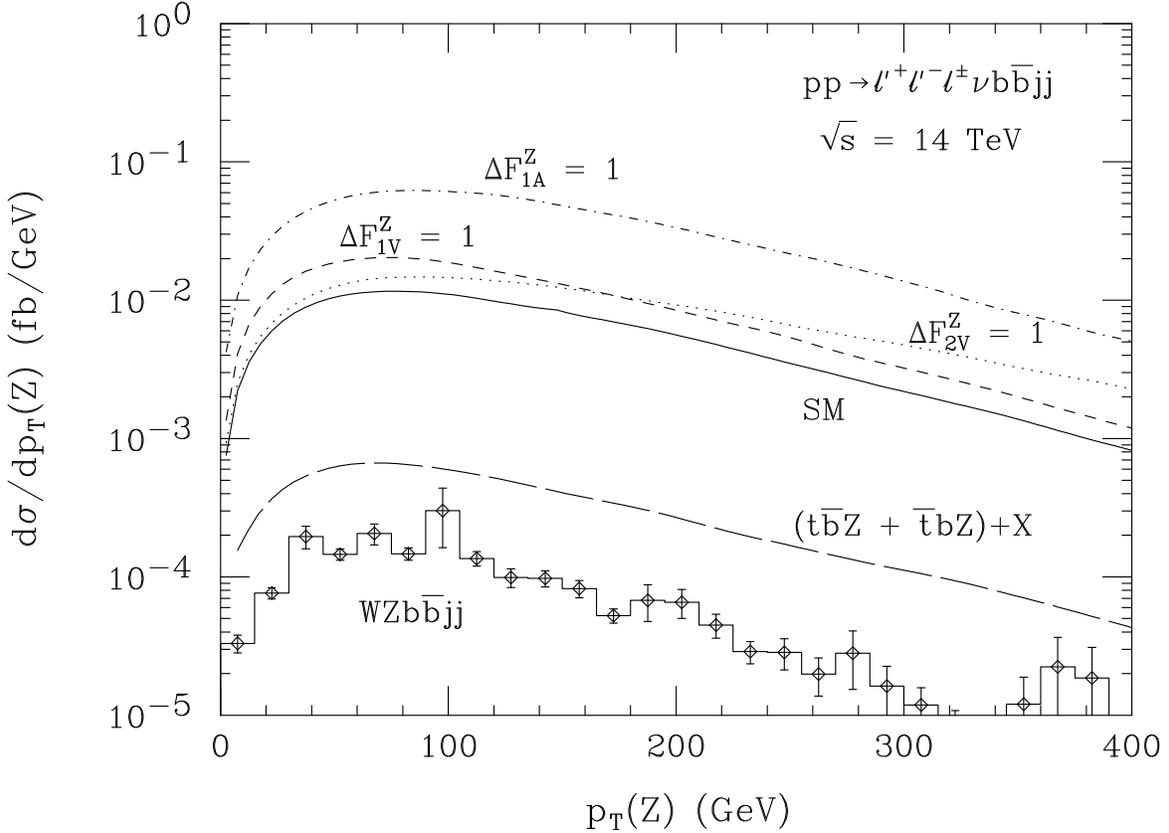} 
\vspace*{2mm}
\caption[]{The differential cross sections at the LHC as a function 
of $p_T(Z)$ for ${\ell'}^+{\ell'}^-\ell\nu b\bar{b}jj$ final states.  
Shown are the SM predictions for $t\bar{t}Z$ production (solid), the 
single-top background (dashed), the $WZb\bar{b}jj$ background 
(histogram), and the predictions for several non-standard $ttZ$ 
couplings.  Only one coupling at a time is allowed to deviate from 
its SM value. The cuts imposed are listed in Eqs.~(\ref{eq:cuts1}), 
(\ref{eq:cuts5}) and~(\ref{eq:cuts4}).}
\label{fig:fig3}
\vspace{-7mm} 
\end{center}
\end{figure}

At the Tevatron, $t\bar{t}Z$ production is quite small, and the
trilepton final state cross section is only about 0.02~fb, far too
small to be observable for the anticipated integrated luminosity in
Run~II.  We therefore consider this signature only for the LHC.  The
$Z$ boson transverse momentum distribution is shown in
Fig.~\ref{fig:fig3} for the SM signal and backgrounds, as well as for 
the signal with several non-standard $ttZ$ couplings.
Only one coupling at a time is allowed to deviate from its SM
prediction.  The backgrounds are each more than one order of magnitude
smaller than the SM signal.  As in $W\gamma b\bar{b}jj$
production, numerical evaluation of the $WZb\bar{b}jj$ helicity
amplitudes is very time consuming.  We thus show its differential
cross section as a histogram, where the error bars represent the Monte
Carlo statistical uncertainty.

Figure~\ref{fig:fig3} shows that, as in the $tt\gamma$ case, the
dimension five couplings $F^Z_{2V,A}$ lead to a significantly harder
transverse momentum distribution.  Furthermore, as in the case of
$F^\gamma_{2V}$ and $F^\gamma_{2A}$, almost identical numerical
results for $F^Z_{2V}$ and $F^Z_{2A}$ are found for equal coupling
strengths, and a $CP$-violating asymmetry similar to
$A^\mu_{cut}(p_{Tcut})$ has to be used to discriminate between the
weak magnetic and weak electric dipole form factors.

Varying $F^Z_{1V,A}$ leads mostly to a cross section
normalization change, hardly affecting the shape of the $p_T(Z)$
distribution.  This is because, unlike in the $tt\gamma$ case, there
is no radiative top decay, i.e. no $t\bar{t}$ events where $t\to WZb$.
This implies that, for the cuts we impose, the $p_T(Z)$ distribution
for SM couplings and for $F^Z_{1V,A}=-F^{Z,SM}_{1V,A}$ are almost
degenerate.

\begin{figure}[t!]
\begin{center}
\includegraphics[width=15.5cm]{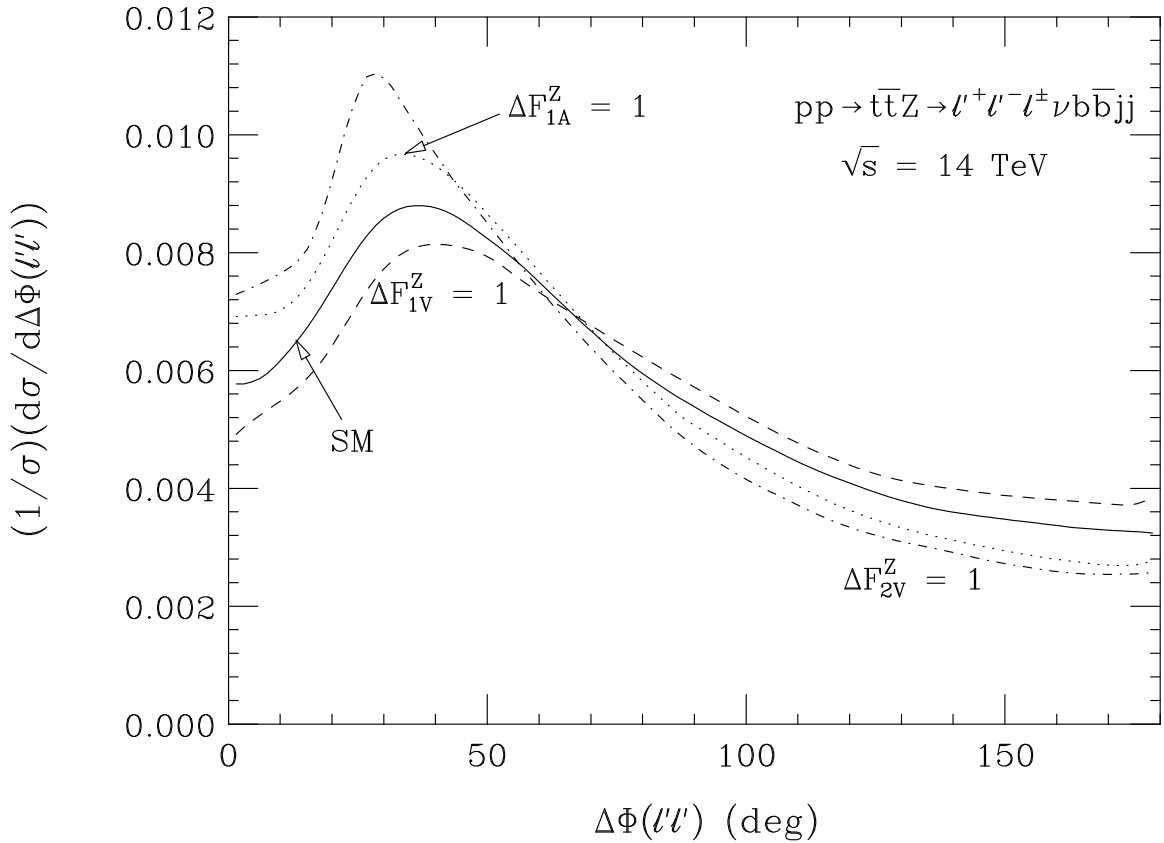} 
\vspace*{2mm}
\caption[]{The normalized differential signal cross sections at LHC 
as a function of the $Z\to{\ell'}^+{\ell'}^-$ azimuthal opening angle,
$\Delta\Phi({\ell'}{\ell'})$.  Shown are the SM distribution (solid
line) and the predictions for several non-standard $ttZ$ couplings.
Only one coupling at a time is allowed
to deviate from its SM prediction.  The cuts imposed are listed in
Eqs.~(\ref{eq:cuts1}), (\ref{eq:cuts5}) and~(\ref{eq:cuts4}).}
\label{fig:fig4}
\vspace{-7mm} 
\end{center}
\end{figure}

Currently, the SM $t\bar{t}Z$ cross section is known only at LO, and
has substantial factorization and renormalization scale uncertainty.
Since the backgrounds are insignificant, this normalization
uncertainty will ultimately be the limiting factor in extracting
anomalous vector and axial vector $ttZ$ couplings, which mostly just
change the normalization. 
To improve sensitivity to $F^Z_{1V,A}$, we need an observable which
changes shape in the presence of anomalous couplings.  An excellent
candidate is the $Z\to\ell'^+\ell'^-$ dilepton azimuthal opening
angle, $\Delta\Phi({\ell'}{\ell'})$.  We show its normalized
distribution for the SM and various anomalous couplings in
Fig.~\ref{fig:fig4}.  
Anomalous vector couplings (dashed line) reduce
the peaking at small opening angles, whereas the opposite is true for
non-standard axial vector couplings (dotted line).  The shape change
is most pronounced for $F^Z_{2V,A}$.  Since the $p_T(Z)$ distribution
is considerably harder in the presence of these couplings, the increased
$Z$ boson Lorentz boost leads to a decrease of
$\Delta\Phi({\ell'}{\ell'})$.

\subsection{The {\boldmath $t\bar{t}Z$} dilepton final state}
\label{sec:sec4b}

\begin{figure}[t!]
\begin{center}
\includegraphics[width=15.5cm]{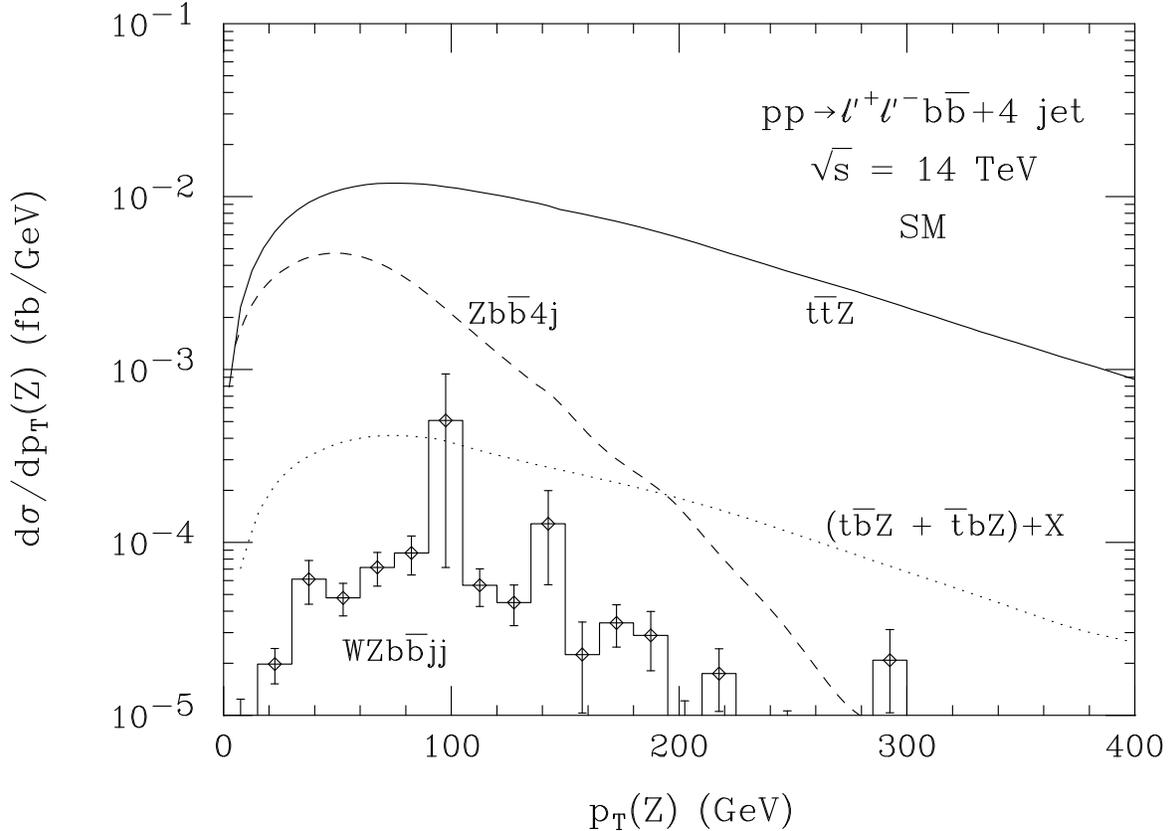} 
\vspace*{2mm}
\caption[]{The differential cross sections at the LHC as a function of 
$p_T(Z)$ for ${\ell'}^+{\ell'}^-b\bar{b}+4j$ final states.  The SM is
the solid curve.  Backgrounds are $Zb\bar{b}+4j$ (dashed), single-top
production (dotted), and $WZb\bar{b}jj$ (histogram).  The cuts imposed
are listed in Eqs.~(\ref{eq:cuts1}) and (\ref{eq:cuts4}-\ref{eq:cuts7}).}
\label{fig:fig5}
\vspace{-7mm} 
\end{center}
\end{figure}

As in the trilepton case, we impose the cuts of Eq.~(\ref{eq:cuts1})
to identify leptons, $b$ quarks and light jets, and again require that
the ${\ell'}^+{\ell'}^-$ invariant mass satisfies
Eq.~(\ref{eq:cuts4}).  The main background arises from
$Zb\bar{b}+4j$ production, which we calculate using {\tt
ALPGEN}~\cite{Mangano:2002ea}.  To adequately suppress it, we
additionally require that events have at least one combination of jets
and $b$ quarks which fulfills the requirements
\begin{eqnarray}\label{eq:cuts6}
m_t - 20~{\rm GeV} < m(b_1j_1j_2) < m_t + 20~{\rm GeV}, & & 
M_W - 20~{\rm GeV} < m(j_1j_2)    < M_W + 20~{\rm GeV}, \\ \label{eq:cuts7}
m_t - 20~{\rm GeV} < m(b_2j_3j_4) < m_t + 20~{\rm GeV}, & & 
M_W - 20~{\rm GeV} < m(j_3j_4)    < M_W + 20~{\rm GeV},
\end{eqnarray}
where $b_{1,2}=b,\bar{b}\,$, and $j_i$, $i=1,\dots, 4$, are the four light
jets.  The SM $p_T(Z)$ distribution, together with those of the
$Zb\bar{b}+4j$, singly-resonant $(t\bar{b}Z+\bar{t}bZ)+X$ and
non-resonant $WZb\bar{b}jj$ backgrounds is shown in
Fig.~\ref{fig:fig5}.
The signatures for anomalous $ttZ$ couplings are similar to those in the
trilepton channel, so we do not show them here.

The non-resonant backgrounds fall much faster with $p_T(Z)$ than the
signal and singly-resonant background.  The $Zb\bar{b}+4j$ background
is important only for $p_T(Z)<100$~GeV.  For $p_T(Z)>200$~GeV,
$(t\bar{b}Z+\bar{t}bZ)+X$ production constitutes the largest
background.  Except for very small values of $p_T(Z)$, the signal to
background ratio (S:B) is significantly better than 1:1.  The SM
signal cross section is approximately the same size as in the dilepton
final state.  We therefore take both channels into account in
extracting anomalous coupling sensitivity limits.  Cross sections for
the signal and backgrounds are summarized in Table~\ref{tab:tab0a}.
\begin{table}[b!]
\caption{Expected LHC cross sections (fb) for the $t\bar{t}Z$ trilepton
and dilepton channels.  The cuts applied are described in the text.  No
particle ID efficiencies are included.}
\label{tab:tab0a}
\vspace{2mm}
\begin{tabular}{ccc}
process & ${\ell'}^+{\ell'}^-\ell\nu b\bar{b}jj$
        & ${\ell'}^+{\ell'}^-b\bar{b}+4j$ \\
\tableline 
signal                    & 2.25 & 2.32 \\
$Zb\bar{b}+4$~jet         &  --  & 0.43 \\
$(t\bar{b}Z+\bar{t}bZ)+X$ & 0.12 & 0.08 \\
$WZb\bar{b}jj$            & 0.03 & 0.02 
\end{tabular}
\end{table}
%

\section{Limits on Anomalous Top Quark Couplings}
\label{sec:sec5}

The shape and normalization changes of the photon or $Z$-boson
transverse momentum distribution and, for $t\bar{t}Z$ production, the
$\Delta\Phi({\ell'}{\ell'})$ distribution, can be used to derive
quantitative sensitivity bounds on the anomalous $tt\gamma$ and $ttZ$
couplings.  We do this by performing a $\chi^2$ test on the
distributions and calculating $68\%$ and $95\%$ confidence level (CL)
limits.  To calculate the statistical significance, we split the
distributions into a number of bins, each with typically more than
five events, approximating the Poisson statistics via a Gaussian
distribution.  We impose the cuts described in Secs.~\ref{sec:sec3}
and~\ref{sec:sec4} and combine channels with electrons and muons in
the final state, conservatively assuming a common lepton
identification efficiency of $\epsilon_\ell=0.85$ for each lepton.  We
take the identification efficiency for photons to be
$\epsilon_\gamma=0.8$ and assume a double $b$-tag efficiency of
$\epsilon_b^2=0.25(0.4)$ at the Tevatron (LHC).  Except for the 
$tt\gamma$ and $ttZ$ couplings we assume the SM to be valid: the $Wtb$ 
and $ttg$ couplings can be precisely measured at the LHC in single
top~\cite{Boos:1999dd} and $t\bar{t}$ production~\cite{Beneke:2000hk}.  
Correlations between different anomalous couplings are fully included.

Our expression for the $\chi^2$ statistics used to compute confidence 
levels is~\cite{babe}
\begin{equation}
\chi^2 = \sum_{i=1}^{n_D} \, \frac{(N_i-fN^0_i)^2}{fN_i^0}+(n_D-1) \, ,
\end{equation}
where $n_D$ is the number of bins, $N_i$ is the number of events for a
given set of anomalous couplings, and $N_i^0$ is the number of events
in the SM in the $i$th bin.  The parameter $f$ reflects the
uncertainty in SM cross section normalization within the allowed
range.  We determine it by minimizing $\chi^2$:
\begin{equation}
f = \begin{cases}
(1+\Delta{\cal N})^{-1} & \text{for $\bar{f} < (1+\Delta{\cal N})^{-1}$} \\
\bar{f} & \text{for $ (1+\Delta{\cal N})^{-1}<\bar{f}<1+\Delta{\cal N}$} \\
1+\Delta{\cal N} & \text{for $\bar{f} > 1+\Delta{\cal N}$} 
\end{cases}
\end{equation}
with
\begin{equation}
\bar{f}^2 = \left\{ \sum_{i=1}^{n_D}N_i^0 \right\}^{-1} \,
\sum_{i=1}^{n_D}\frac{N_i^2}{N_i^0} \; .
\end{equation}
The parameter $\Delta{\cal N}$ is the SM cross section uncertainty.
It arises primarily from the currently-unknown signal QCD corrections,
and from PDF uncertainties.  In the following we assume $\Delta{\cal
N}=30\%$ unless stated otherwise.  We universally assume real
anomalous couplings.

\subsection{Sensitivity bounds for {\boldmath $tt\gamma$} couplings}
\label{sec:sec5a}

To derive sensitivity bounds for anomalous $tt\gamma$ couplings, we
take into account the $t\bar{t}j$, singly-resonant
$(t\bar{b}\gamma+\bar{t}b\gamma)+X$, and $W\gamma b\bar{b}jj$
backgrounds.  The variation of the singly-resonant background with
$tt\gamma$ anomalous couplings is ignored.  For the probabilities that 
a jet fakes a photon at the Tevatron and LHC we use the values listed 
in Sec.~\ref{sec:sec3b}.  All other backgrounds are assumed to be
negligible.  For the Tevatron, we derive sensitivity limits for an
integrated luminosity of 8~fb$^{-1}$ which is the total integrated
luminosity anticipated for Run~II.  For the LHC we calculate bounds
for 30~fb$^{-1}$, 300~fb$^{-1}$, and 3000~fb$^{-1}$. An integrated
luminosity of 300~fb$^{-1}$ corresponds to 3~years of running at the
LHC design luminosity of ${\cal L}=10^{34}\,{\rm cm}^{-2}\,s^{-1}$.
The smaller value of 30~fb$^{-1}$ is expected for the first few years
of operation of the LHC when the luminosity is likely to be
significantly smaller than design. The larger value of 3000~fb$^{-1}$
can be achieved in about 3~years of running at a luminosity-upgraded
LHC.

\begin{figure}[t!]
\begin{center}
\includegraphics[width=12.5cm]{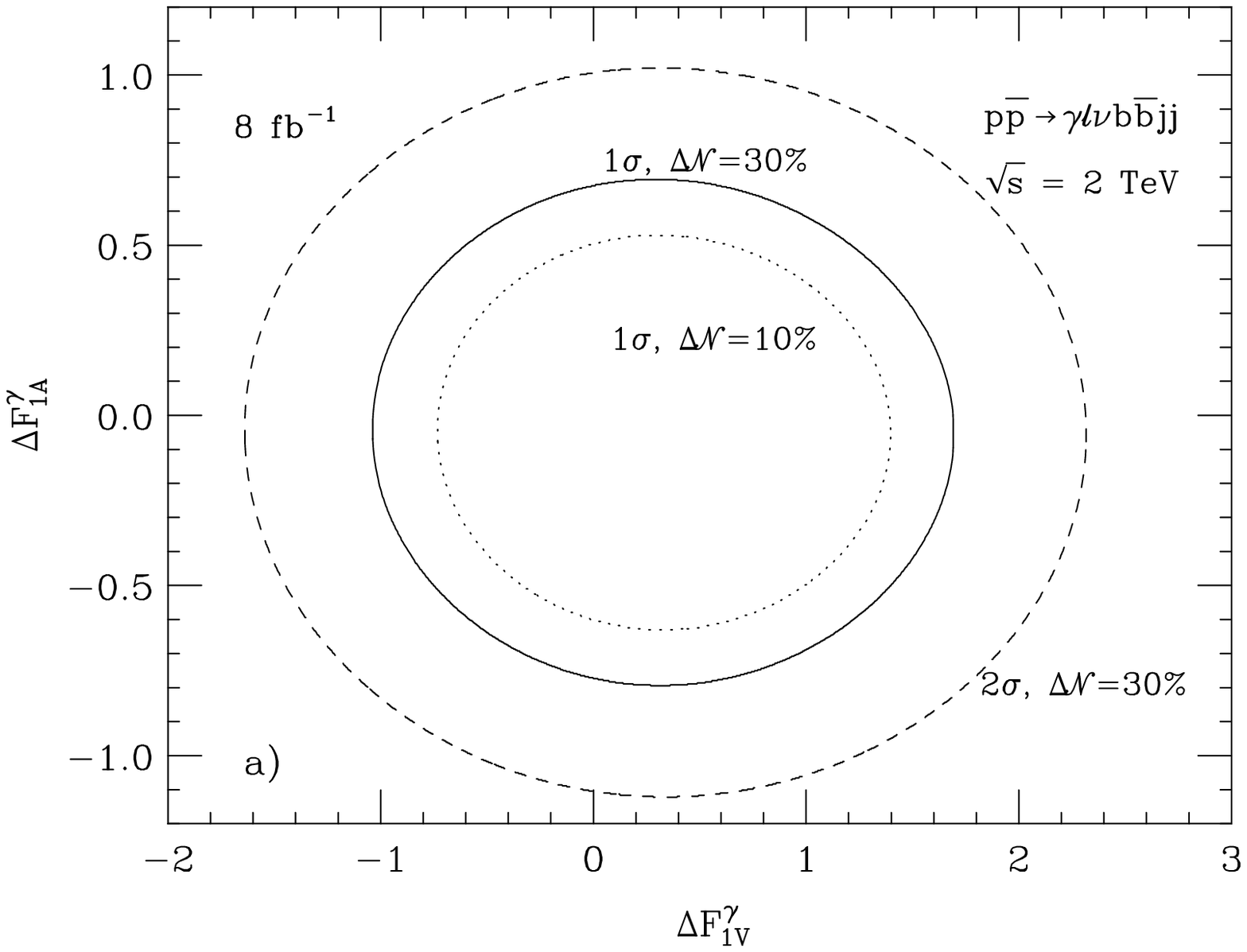} \\
\includegraphics[width=12.8cm]{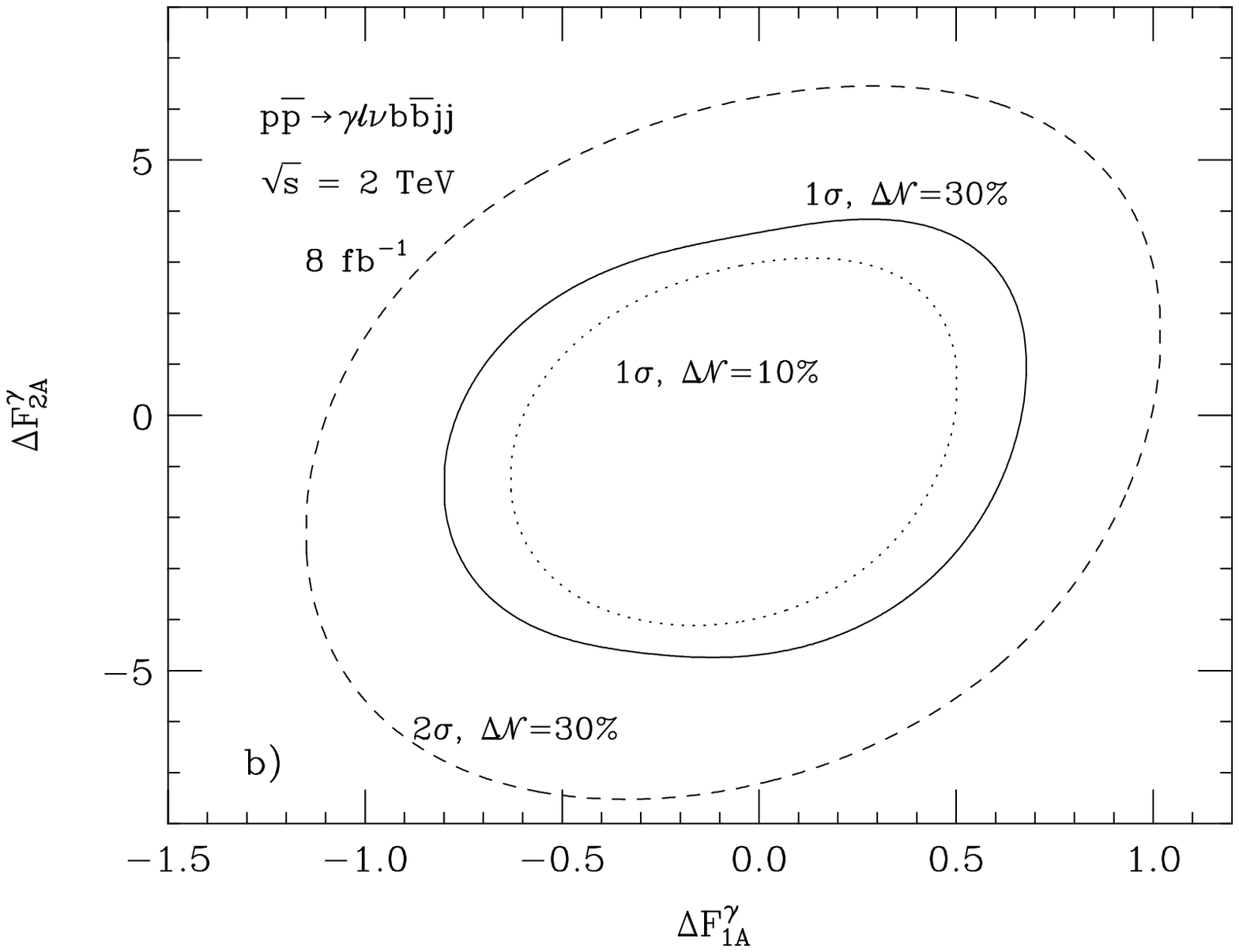} 
\vspace*{2mm}
\caption[]{Projected bounds on anomalous $tt\gamma$ couplings for 
$p\bar{p}\to\gamma\ell\nu_\ell b\bar{b}jj$ at the Tevatron and an
integrated luminosity of 8~fb$^{-1}$.  Shown are $68.3\%$ (solid) and
$95\%$ CL limits (dashed) for a SM cross section normalization
uncertainty of $\Delta {\cal N}=30\%$, and the $68.3\%$ CL limits for
$\Delta {\cal N}=10\%$: (a) for $\Delta F^\gamma_{1A}$ versus $\Delta
F^\gamma_{1V}$; and (b) for $\Delta F^\gamma_{2A}$ versus $\Delta
F^\gamma_{1A}$.  In each graph, only those couplings which are plotted
against each other are assumed to be different from their SM values.}
\label{fig:fig6}
\vspace{-7mm}  
\end{center}
\end{figure}

Our results for the Tevatron are shown in Table~\ref{tab:tab1}. 
\begin{table}
\caption{Sensitivities achievable at $68.3\%$ and $95\%$ CL for 
anomalous $tt\gamma$ couplings in $p\bar{p}\to\gamma\ell\nu_\ell
b\bar{b}jj$ at the Tevatron ($\sqrt{s}=2$~TeV) for an integrated
luminosity of 8~fb$^{-1}$.  The limits shown represent the maximum and
minimum values obtained when taking into account the correlations
between any possible pair of anomalous couplings. The cuts imposed are
described in Sec.~\ref{sec:sec3a}.}
\label{tab:tab1}
\vspace{2mm}
\begin{tabular}{ccc}
coupling & $68.3\%$ CL & $95\%$ CL \\
\tableline 
$\Delta F^\gamma_{1V}$ & $\begin{matrix} +1.92 \\[-4pt]
-1.20\end{matrix}$ & $\begin{matrix} +2.60 \\[-4pt]
-1.88\end{matrix}$ \\
$\Delta F^\gamma_{1A}$ & $\begin{matrix} +0.69 \\[-4pt]
-0.82\end{matrix}$ & $\begin{matrix} +1.03 \\[-4pt]
-1.17\end{matrix}$ \\
$\Delta F^\gamma_{2V}$ & $\begin{matrix} +5.16 \\[-4pt]
-5.21\end{matrix}$ & $\begin{matrix} +8.49 \\[-4pt]
-8.73\end{matrix}$ \\
$\Delta F^\gamma_{2A}$ & $\begin{matrix} +5.19 \\[-4pt]
-5.08\end{matrix}$ & $\begin{matrix} +7.85 \\[-4pt]
-8.43\end{matrix}$           
\end{tabular}
\end{table}
The correlations between various anomalous $tt\gamma$ couplings are
illustrated in Fig.~\ref{fig:fig6} for two combinations, $\Delta
F^\gamma_{1V}$ versus $\Delta F^\gamma_{1A}$, and $\Delta
F^\gamma_{1A}$ versus $\Delta F^\gamma_{2A}$. Correlations between the
couplings are seen to be fairly small at Tevatron energies. This is
also the case for the combinations not shown.

Due to the small cross section and the complicating ``background''
from photon radiation off initial state quarks, Tevatron experiments
are essentially insensitive to the dipole form factors
$F^\gamma_{2V,A}$.  The achievable bounds for these are worse than the
limits from $S$-matrix unitarity for a form factor scale
$\Lambda_{FF}\geq 1$~TeV.  However, for the $tt\gamma$ vector and
axial vector couplings, which are not (directly or indirectly)
constrained by any existing experiment, CDF and D\O\ will be able to
perform a first, albeit not very precise, measurement.  The prospects
are most favorable for $F^\gamma_{1A}$, which, as shown in
Table~\ref{tab:tab1}, can be determined with an accuracy of about
$70\%$ for a SM cross section normalization uncertainty of $30\%$.

As shown in Fig.~\ref{fig:fig6}a, the precision on $F^\gamma_{1A}$ can
be improved to about $50\%$ if the normalization uncertainty can be
reduced to $10\%$.  This depends critically on the signal
normalization.  Currently, the $t\bar{t}\gamma$ cross section is known
only at LO.  Once the NLO QCD corrections are known, a $10\%$
normalization uncertainty may be realistic.  

The bound on $F^\gamma_{1A}$ can, in principle, be further tightened
by enlarging the signal sample by requiring only one $b$-tagged jet.
As mentioned before, the increase in signal statistics when including
single tagged events comes at the price of increased background.  To
quantify the improvement, detailed simulations are needed.

\begin{figure}[t!]
\begin{center}
\begin{tabular}{lr}
\includegraphics[width=8.2cm]{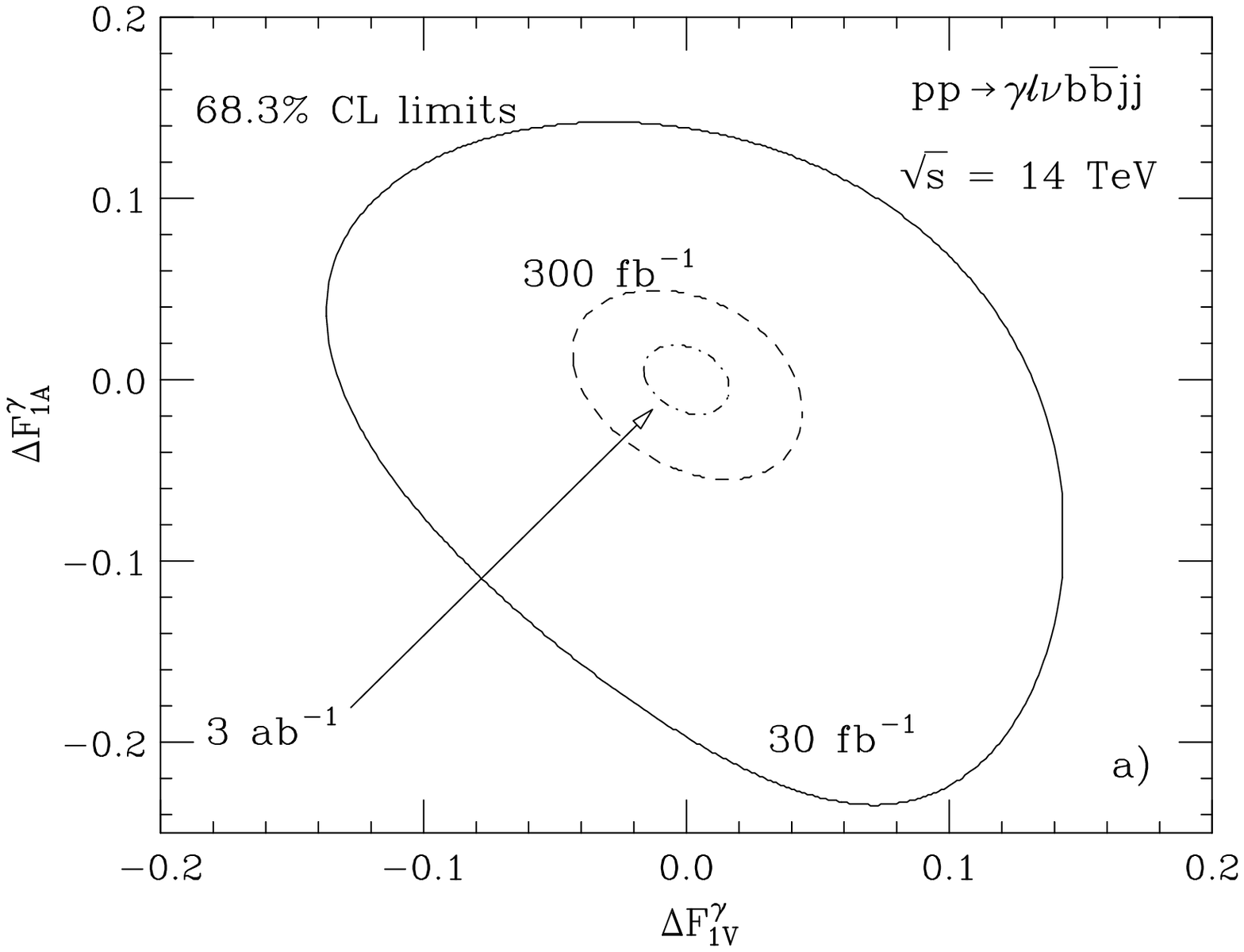} &
\includegraphics[width=8.2cm]{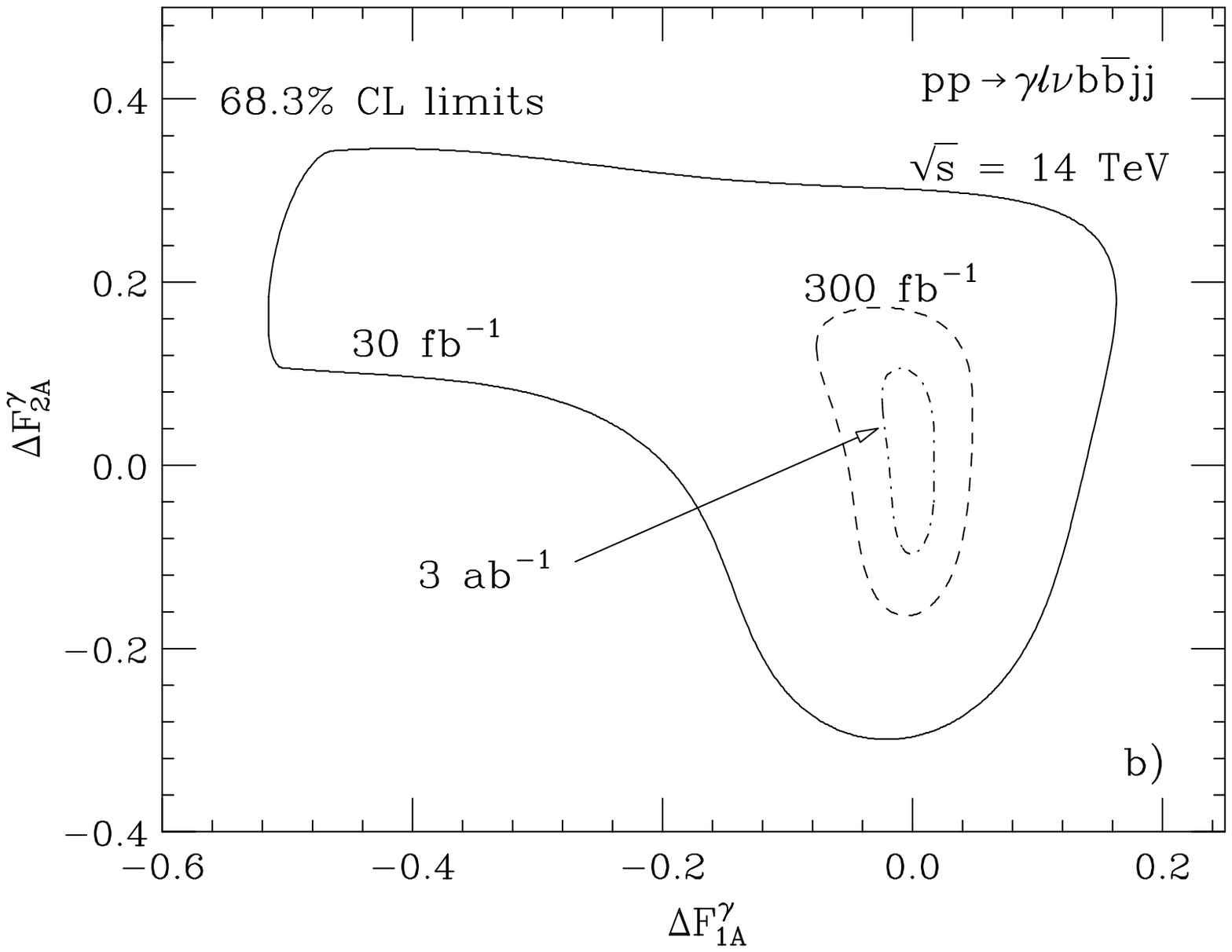} \\
\includegraphics[width=8.2cm]{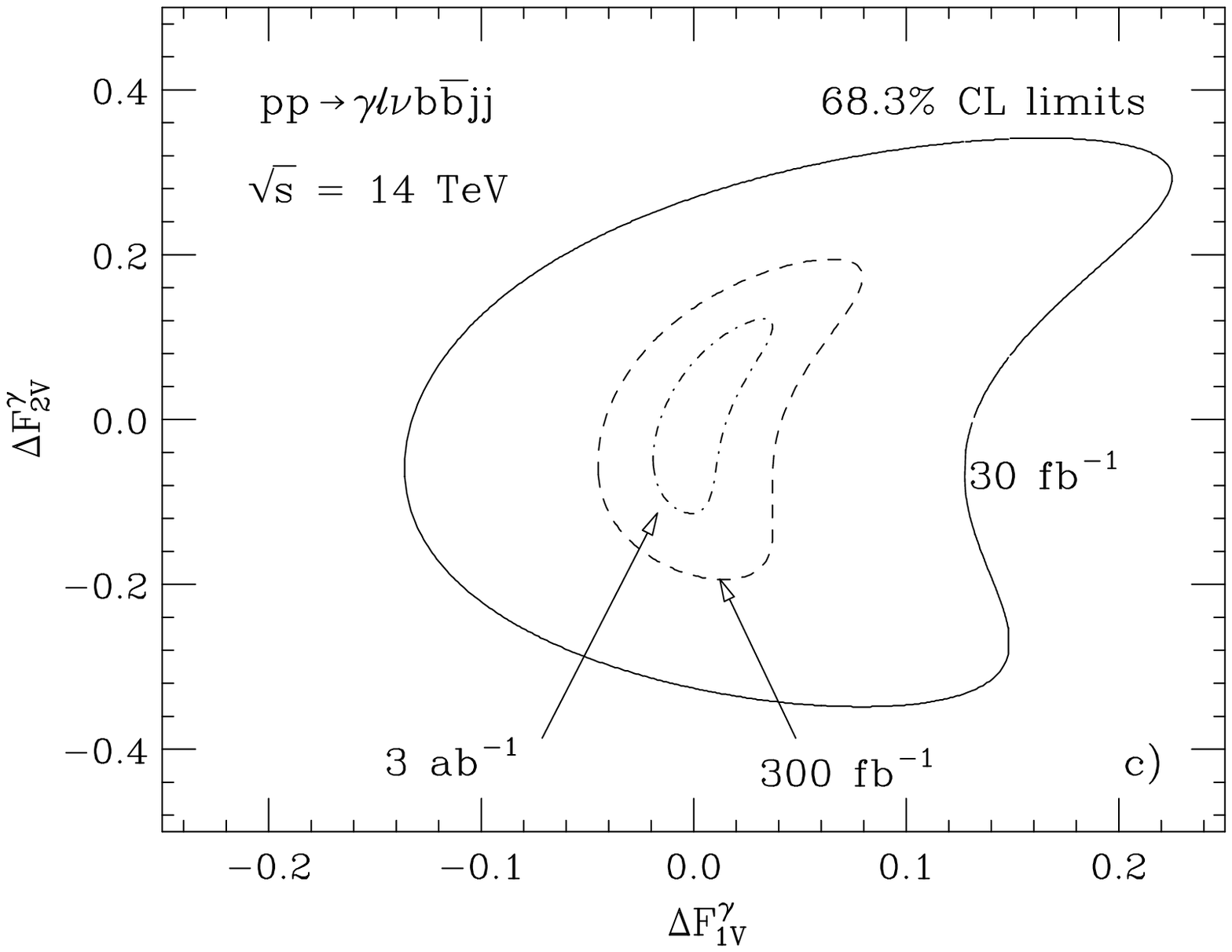} &
\includegraphics[width=8.2cm]{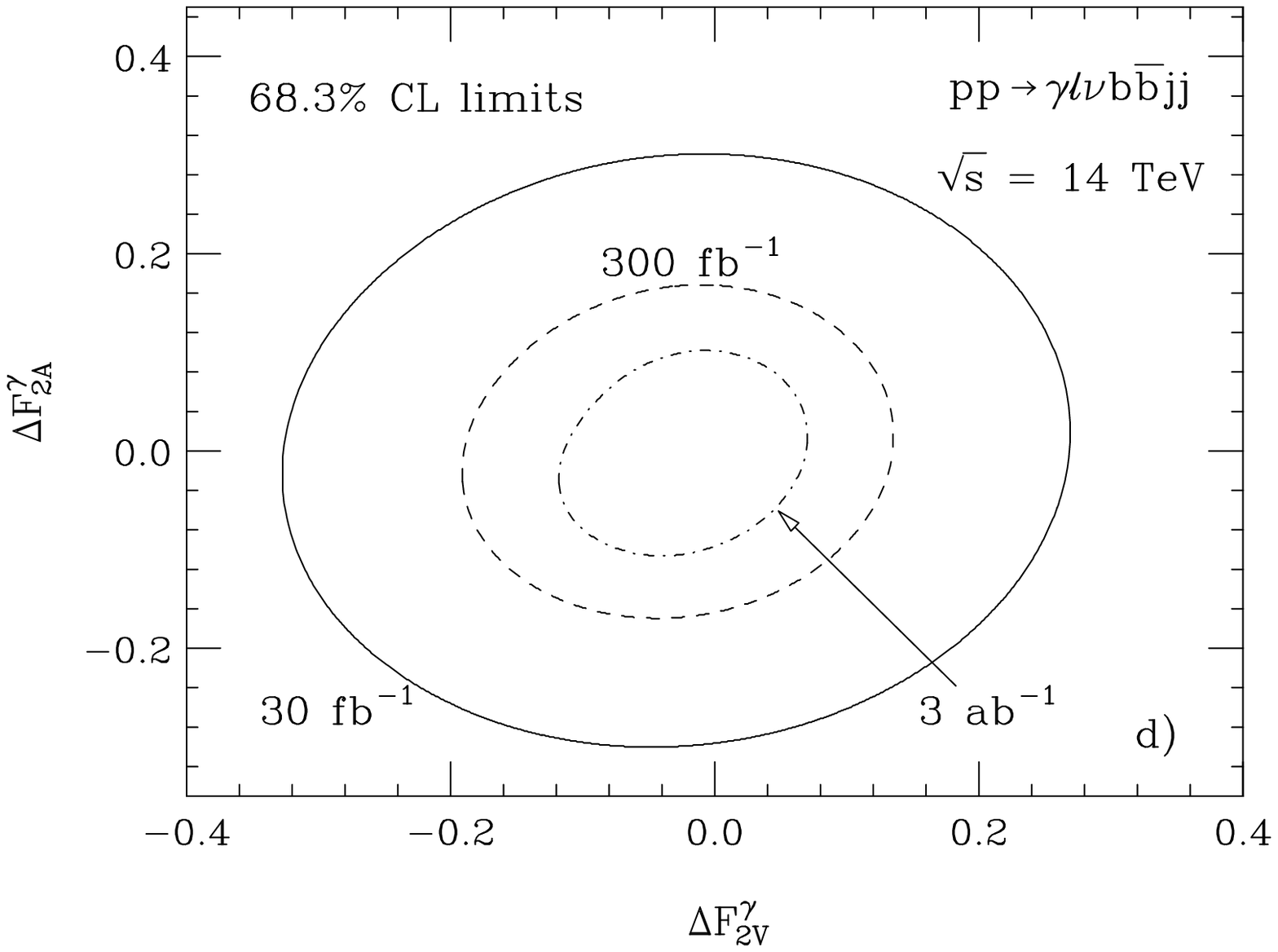} 
\end{tabular}
\vspace*{2mm}
\caption[]{Projected bounds on anomalous $tt\gamma$ couplings for 
$pp\to\gamma\ell\nu_\ell b\bar{b}jj$ at the LHC.  Shown are $68.3\%$
CL limits for a SM cross section normalization uncertainty of $\Delta
{\cal N}=30\%$ and for integrated luminosities of 30~fb$^{-1}$
(solid), 300~fb$^{-1}$ (dashed), and 3000~fb$^{-1}$ (dot-dashed): (a)
for $\Delta F^\gamma_{1A}$ versus $\Delta F_{1V}^\gamma$, (b) for
$\Delta F^\gamma_{2A}$ versus $\Delta F^\gamma_{1A}$, (c) for $\Delta
F^\gamma_{2V}$ versus $\Delta F^\gamma_{1V}$ and (d) for $\Delta
F^\gamma_{2A}$ versus $\Delta F^\gamma_{2V}$.  In each graph, only
those couplings which are plotted against each other are assumed to be
different from their SM values.}
\label{fig:fig7}
\vspace{-7mm}  
\end{center}
\end{figure}

The sensitivity bounds achievable at the LHC are shown in
Table~\ref{tab:tab2} and Fig.~\ref{fig:fig7}.
\begin{table}[t!]
\caption{Sensitivities achievable at $68.3\%$ CL for anomalous
$tt\gamma$ couplings in $pp\to \gamma\ell\nu_\ell b\bar{b}jj$ at the
LHC ($\sqrt{s} = 14$~TeV) for an integrated luminosities of
30~fb$^{-1}$, 300~fb$^{-1}$, and 3000~fb$^{-1}$.  The limits shown
represent the maximum and 
minimum values obtained when taking into account the correlations
between any possible pair of anomalous couplings. The cuts
imposed are described in Sec.~\ref{sec:sec3a}.}
\label{tab:tab2}
\vspace{2mm}
\begin{tabular}{cccc}
coupling & 30~fb$^{-1}$ & 300~fb$^{-1}$ & 3000~fb$^{-1}$ \\
\tableline 
$\Delta F^\gamma_{1V}$ & $\begin{matrix} +0.23 \\[-4pt]
-0.14\end{matrix}$ & $\begin{matrix} +0.079 \\[-4pt]
-0.045\end{matrix}$ & $\begin{matrix} +0.037 \\[-4pt]
-0.019\end{matrix}$ \\
$\Delta F^\gamma_{1A}$ & $\begin{matrix} +0.17 \\[-4pt]
-0.52\end{matrix}$ & $\begin{matrix} +0.051 \\[-4pt]
-0.077\end{matrix}$ & $\begin{matrix} +0.018 \\[-4pt]
-0.024\end{matrix}$ \\
$\Delta F^\gamma_{2V}$ & $\begin{matrix} +0.34 \\[-4pt]
-0.35\end{matrix}$ & $\begin{matrix} +0.19 \\[-4pt]
-0.20\end{matrix}$ & $\begin{matrix} +0.12 \\[-4pt]
-0.12\end{matrix}$ \\
$\Delta F^\gamma_{2A}$ & $\begin{matrix} +0.35 \\[-4pt]
-0.36\end{matrix}$ & $\begin{matrix} +0.19 \\[-4pt]
-0.21\end{matrix}$  & $\begin{matrix} +0.11 \\[-4pt]
-0.14\end{matrix}$          
\end{tabular}
\end{table}
Even for a modest integrated luminosity of 30~fb$^{-1}$, one expects
more than 500~signal events after acceptances and efficiencies are
taken into account. This will make it possible to measure the
$tt\gamma$ vector and axial vector couplings, and the dipole form
factors, with a precision of typically $20\%$ and $35\%$,
respectively. For 300~fb$^{-1}$, the limits improve to $4-7\%$ for
$F^\gamma_{1V,A}$ and to about $20\%$ for $F^\gamma_{2V,A}$. At the
SLHC, assuming an integrated luminosity of 3000~fb$^{-1}$, one can
hope to achieve a $2-3\%$ measurement of the vector and axial vector
couplings, and a $10\%$ measurement of $F^\gamma_{2V,A}$, provided
that particle identification efficiencies are not substantially
smaller, and the reducible backgrounds not much larger, than what we
have assumed.

As shown in Fig.~\ref{fig:fig7}, with the exception of $F^\gamma_{2V}$
and $F^\gamma_{2A}$ (see Fig.~\ref{fig:fig7}d), there are substantial
correlations between the $tt\gamma$ couplings at the LHC, in
particular for low integrated luminosities where small changes in the
shape of the $p_T(\gamma)$ distribution are not resolved with the
available statistics. Varying only one coupling at a time thus will
produce overly optimistic limits. The correlations between
$F^\gamma_{2V}$ and $F^\gamma_{1A}$ ($F^\gamma_{2A}$ and
$F^\gamma_{1V}$) are similar to those for $F^\gamma_{2A}$ and
$F^\gamma_{1A}$ ($F^\gamma_{2V}$ and $F^\gamma_{1V}$) and thus not
shown in Fig.~\ref{fig:fig7}.

\subsection{Sensitivity bounds for $ttZ$ couplings}
\label{sec:sec5b}

To extract bounds on the $ttZ$ couplings, we perform a simultaneous
fit to the $p_T(Z)$ and the $\Delta\Phi({\ell'}{\ell'})$
distributions, using both the trilepton and dilepton final states.
Since the $(t\bar{b}Z+\bar{t}bZ)+X$ and $WZb\bar{b}jj$ backgrounds are
very small, we take only the $Zb\bar{b}+4j$ background into account in
our $\chi^2$ analysis.  We calculate sensitivity bounds for
300~fb$^{-1}$ and 3000~fb$^{-1}$ at the LHC; for 30~fb$^{-1}$ the
number of events expected is too small to yield meaningful results.

\begin{figure}[t!]
\begin{center}
\begin{tabular}{lr}
\includegraphics[width=8.2cm]{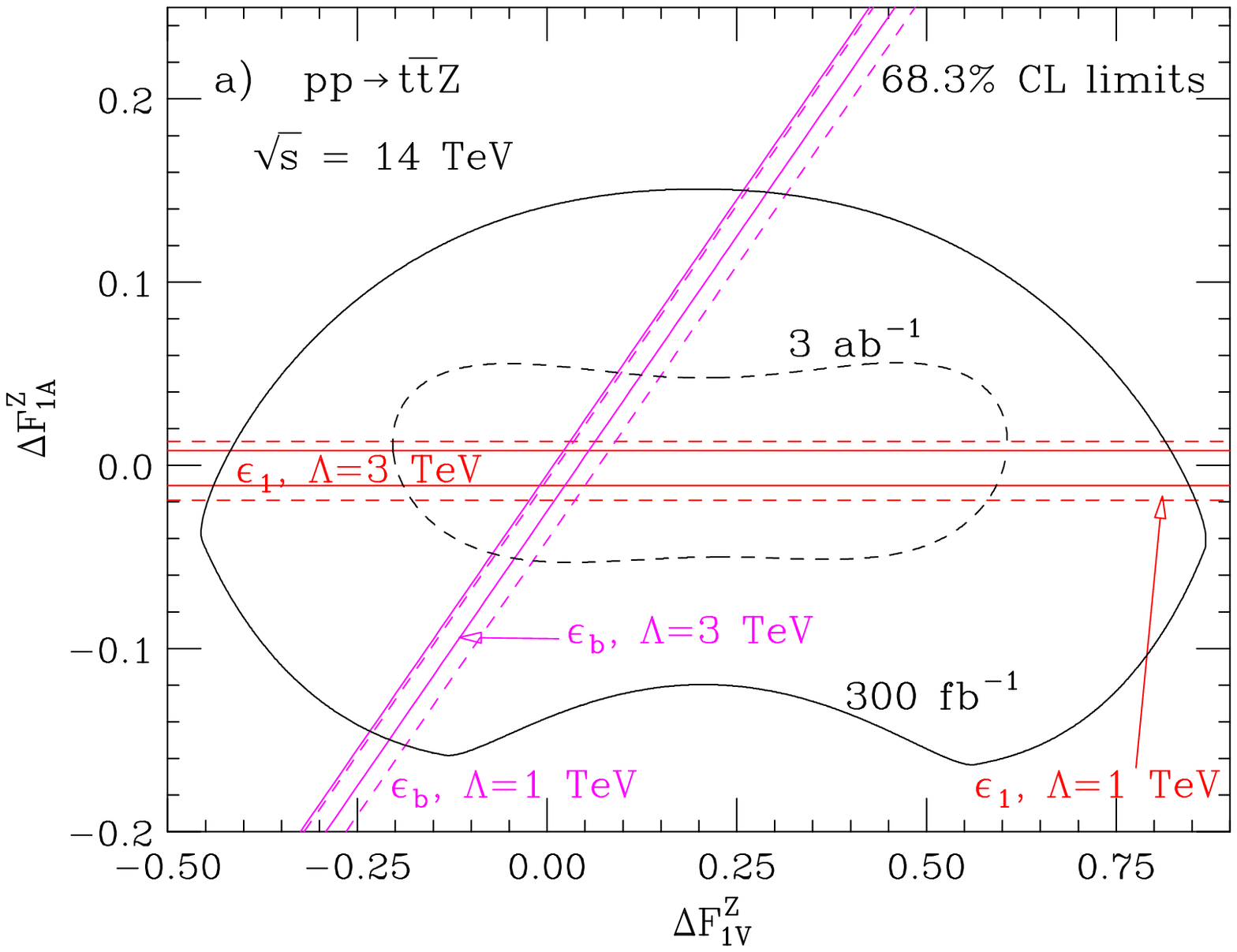} &
\includegraphics[width=8.2cm]{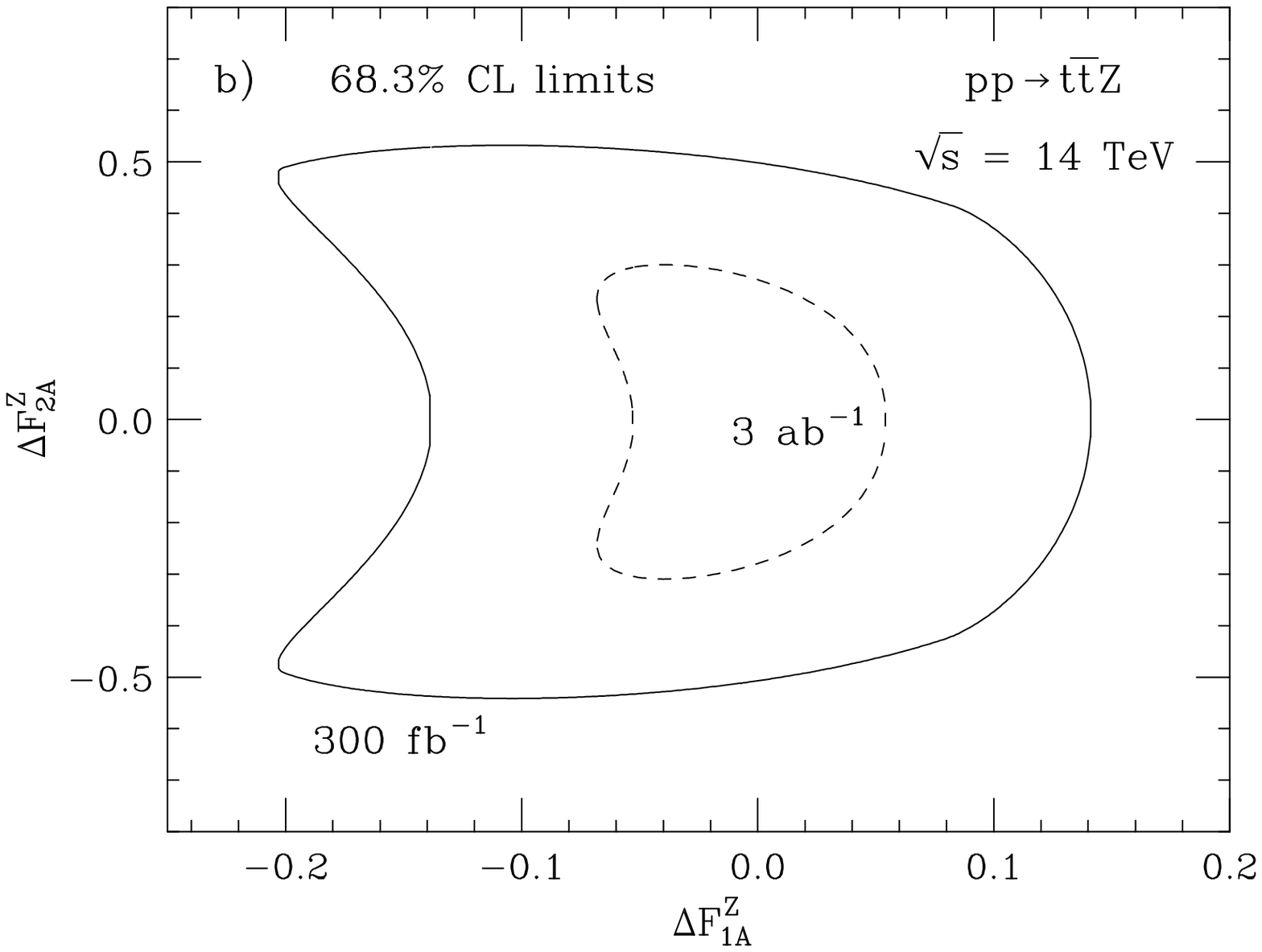} \\
\includegraphics[width=8.2cm]{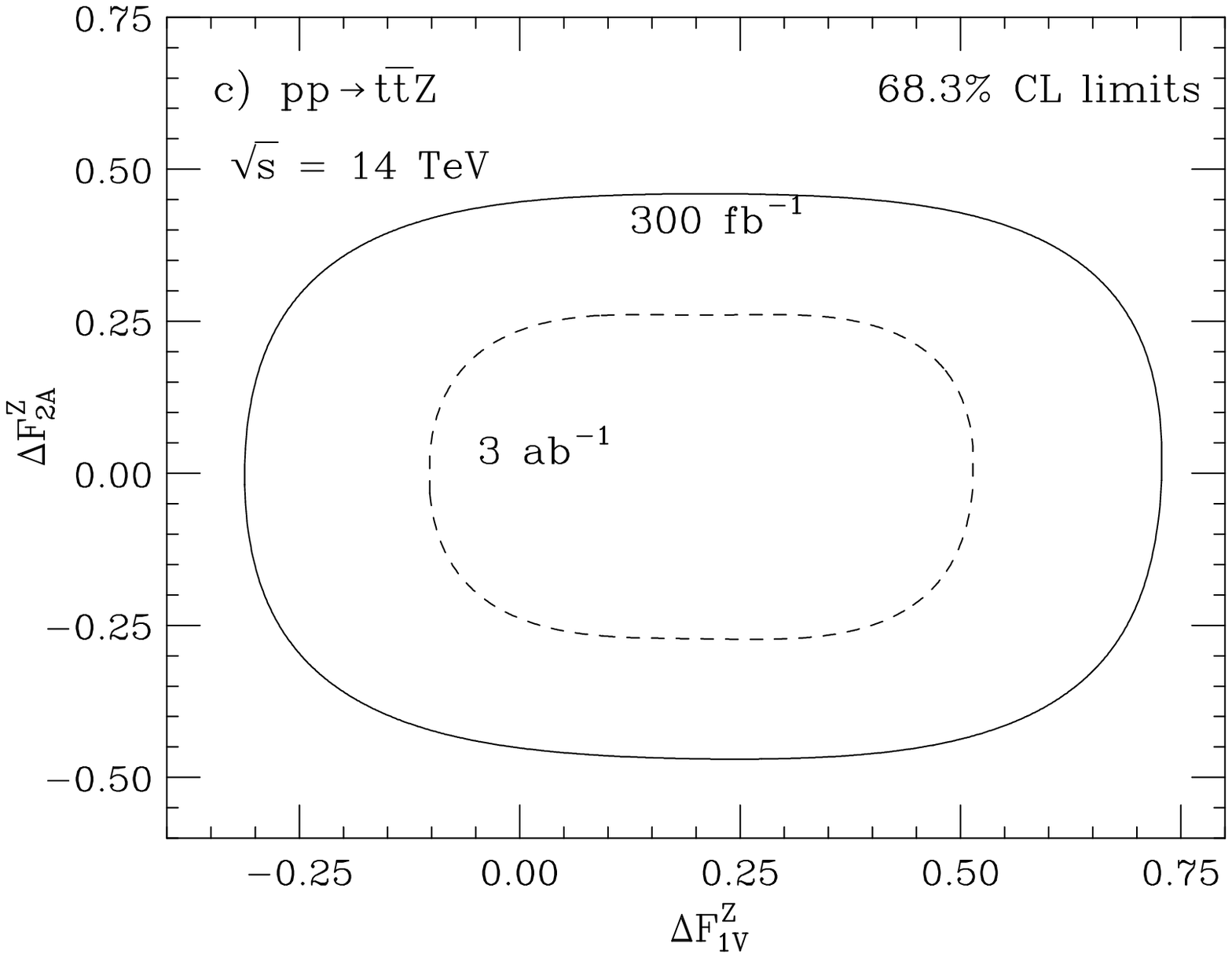} &
\includegraphics[width=8.2cm]{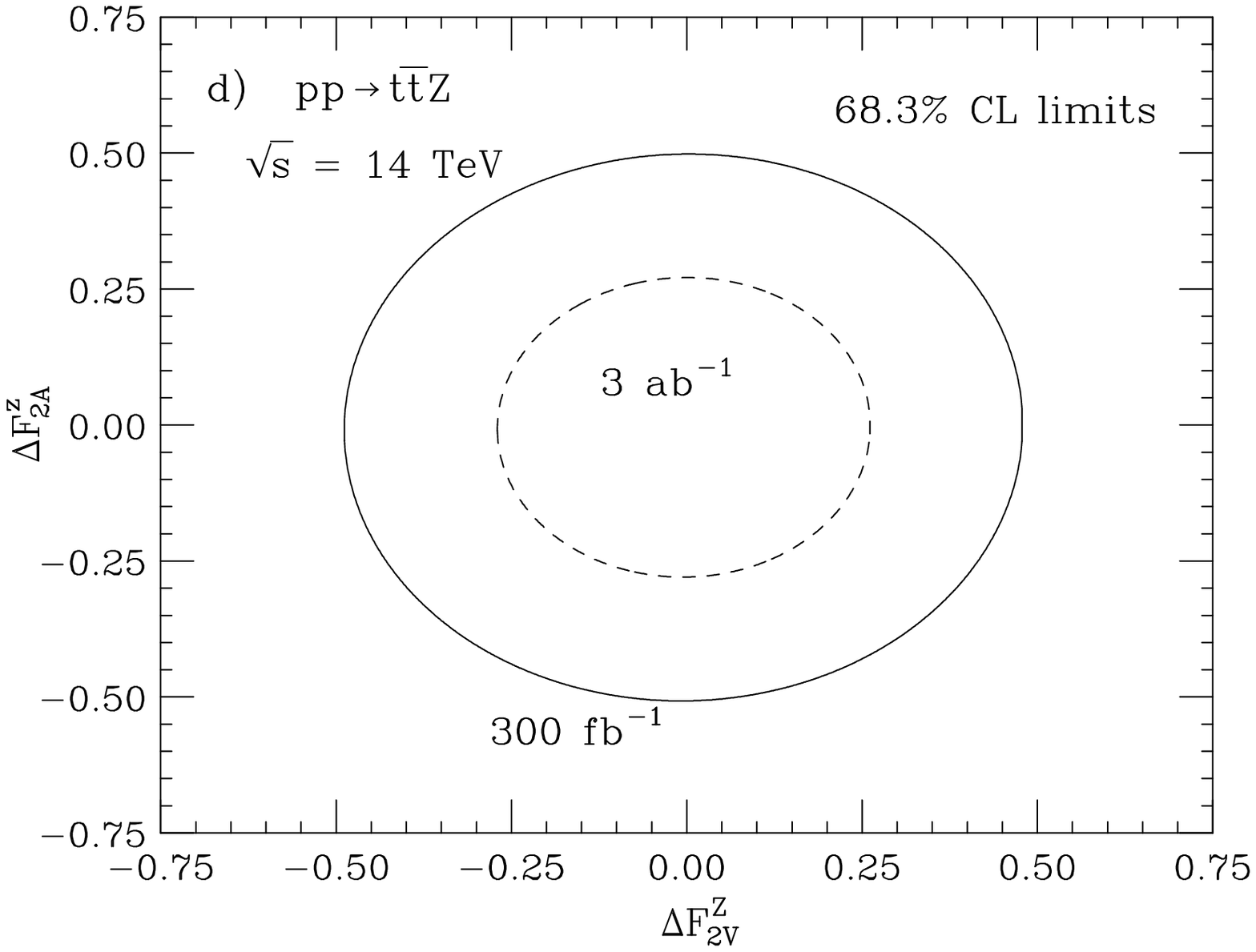} 
\end{tabular}
\vspace*{2mm}
\caption[]{Projected bounds on anomalous $ttZ$ couplings for at the LHC.
Shown are $68.3\%$ CL limits for integrated luminosities of
300~fb$^{-1}$ (solid) and 3000~fb$^{-1}$ (dashed): (a) for $\Delta
F^Z_{1A}$ versus $\Delta F^Z_{1V}$, (b) for $\Delta F^Z_{2A}$ versus
$\Delta F^Z_{1A}$, (c) for $\Delta F^Z_{2V}$ versus $\Delta F^Z_{1V}$
and (d) for $\Delta F^Z_{2A}$ versus $\Delta F^Z_{2V}$.  In (a) we
also include the (indirect) constraints from LEP data (see
Eqs.~(\ref{eq:eps1}) and~(\ref{eq:eps2})) for two choices of the loop
momentum cutoff scale $\Lambda$ (solid: $\Lambda=3$~TeV, dashed:
$\Lambda=1$~TeV).  In each graph, only those couplings which are
plotted against each other are assumed to be different from their SM
values.}
\label{fig:fig8}
\vspace{-7mm}  
\end{center}
\end{figure}

Our results are shown in Table~\ref{tab:tab3} and Fig.~\ref{fig:fig8}.
\begin{table}
\caption{Sensitivities achievable at $68.3\%$ CL for anomalous $ttZ$ 
couplings at the LHC for integrated luminosities of 300~fb$^{-1}$, and
3000~fb$^{-1}$.  The limits shown represent the maximum and 
minimum values obtained when taking into account the correlations
between any possible pair of anomalous couplings. The cuts imposed are
described in Secs.~\ref{sec:sec3a} and~\ref{sec:sec4a}.}
\label{tab:tab3}
\vspace{2mm}
\begin{tabular}{ccc}
coupling & 300~fb$^{-1}$ & 3000~fb$^{-1}$ \\
\tableline 
$\Delta F^Z_{1V}$   & $\begin{matrix} +0.87  \\[-4pt]
-0.46\end{matrix}$  & $\begin{matrix} +0.62  \\[-4pt]
-0.22\end{matrix}$ \\
$\Delta F^Z_{1A}$   & $\begin{matrix} +0.15 \\[-4pt]
-0.20\end{matrix}$ & $\begin{matrix} +0.056 \\[-4pt]
-0.074\end{matrix}$ \\
$\Delta F^Z_{2V}$   & $\begin{matrix} +0.52  \\[-4pt]
-0.52\end{matrix}$  & $\begin{matrix} +0.30  \\[-4pt]
-0.29\end{matrix}$ \\
$\Delta F^Z_{2A}$   & $\begin{matrix} +0.54  \\[-4pt]
-0.53\end{matrix}$  & $\begin{matrix} +0.30  \\[-4pt]
-0.31\end{matrix}$          
\end{tabular}
\end{table}
\renewcommand{\bottomfraction}{0.9}
\renewcommand{\textfraction}{0}
For an integrated luminosity of 300~fb$^{-1}$, it will be possible to
measure the $ttZ$ axial vector coupling with a precision of $15-20\%$,
and $F^Z_{2V,A}$ with a precision of $50-55\%$.  At the SLHC, these
bounds can be improved by factors of about~1.6 ($F^Z_{2V,A}$)
and~3 ($F^Z_{1A}$).  The bounds which can be achieved for
$F^Z_{1V}$ are much weaker than those projected for $F^Z_{1A}$.  As
mentioned in Sec.~\ref{sec:sec4b}, the $p_T(Z)$ distributions for the
SM and for $F^Z_{1V,A}=-F^{Z,SM}_{1V,A}$ are almost degenerate.
This is also the case for the $\Delta\Phi({\ell'}{\ell'})$
distribution.  In a fit to these two distributions, therefore, an area
centered at $\Delta F^Z_{1V,A}=-2F^{Z,SM}_{1V,A}$ remains which cannot
be excluded, even at the SLHC where one expects several thousand
$t\bar{t}Z$ events.  For $F^Z_{1V}$, the two regions merge, resulting
in rather poor limits.  For $F^Z_{1A}$, the two regions are distinct.
Since the area centered at $\Delta F^Z_{1A}=-2F^{Z,SM}_{1A}$ is 
incompatible with the indirect limits on the $ttZ$ vector and axial 
vector couplings from LEP data, it is not included in 
Table~\ref{tab:tab3} or Fig.~\ref{fig:fig8}.

The bounds on the $ttZ$ couplings, with the exception of $F^Z_{2V}$
and $F^Z_{2A}$ (see Fig.~\ref{fig:fig8}d), are strongly
correlated.  The correlations between $F^Z_{2V}$ and $F^Z_{1A}$
($F^Z_{2A}$ and $F^Z_{1V}$) are similar to those for $F^Z_{2A}$ and
$F^Z_{1A}$ ($F^Z_{2V}$ and $F^Z_{1V}$) and thus are not shown. In
Fig.~\ref{fig:fig8}a we also include the indirect bounds resulting
from LEP data (see Eqs.~(\ref{eq:eps1}) and~(\ref{eq:eps2})) for two
choices of the loop momentum cutoff scale $\Lambda$.

\begin{figure}[t!]
\begin{center}
\includegraphics[width=14.5cm]{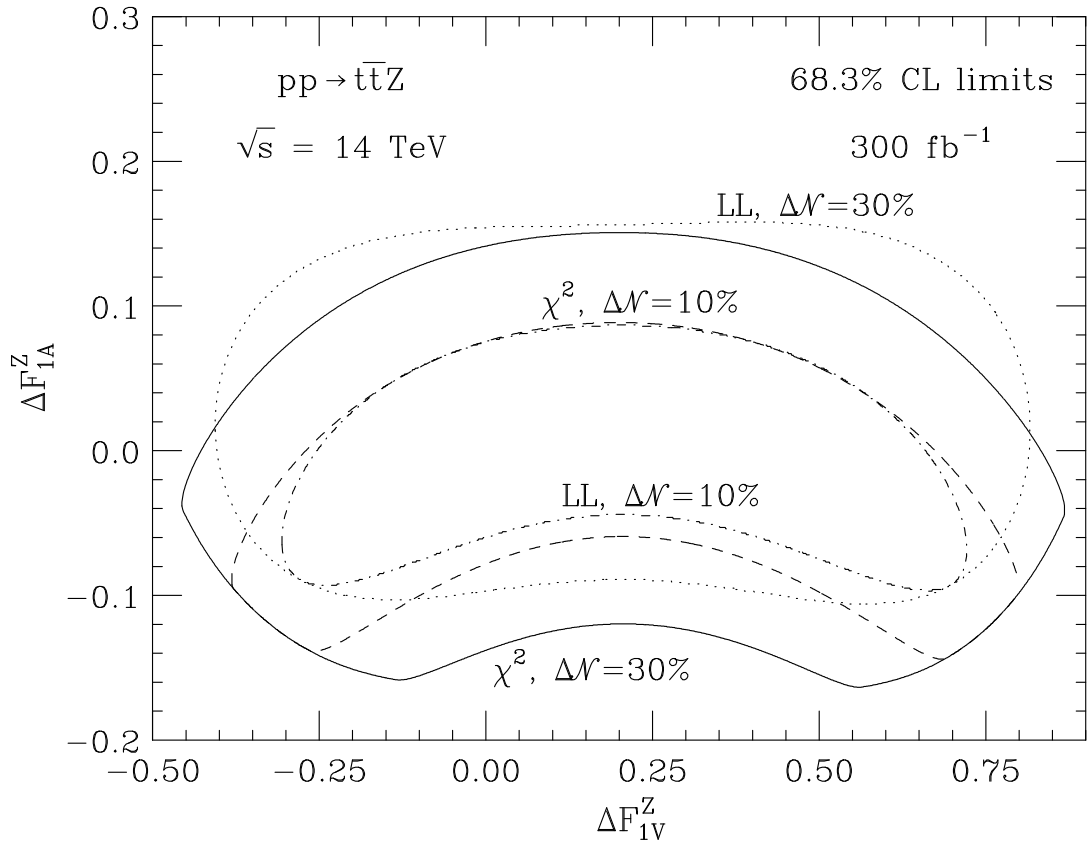} 
\vspace*{2mm}
\caption[]{Projected $68.3\%$ CL bounds on $\Delta F^Z_{1A}$ and 
$\Delta F^Z_{1V}$ for 300~fb$^{-1}$ at the LHC.  The solid and dashed 
curves show the sensitivity bounds obtained using the $\chi^2$ test 
described at the beginning of this section for $\Delta{\cal N}=30\%$ 
and $\Delta{\cal N}=10\%$, respectively.  The dotted and dot-dashed 
lines correspond to the limits found using the log-likelihood method. 
All other $ttZ$ couplings are assumed to have their SM values. }
\label{fig:fig9}
\vspace{-7mm} 
\end{center}
\end{figure}

To test the robustness of our sensitivity limits for anomalous $ttZ$ 
couplings, we have performed an independent analysis using Poisson
statistics and the log-likelihood method.  The normalization 
uncertainty in this approach is treated as a Gaussian fluctuation 
with standard deviation $\Delta{\cal N}$.  Except for $F^Z_{1A}$, 
the limits obtained using the log-likelihood method are similar to 
those shown in Table~\ref{tab:tab3} and Fig.~\ref{fig:fig8}; they 
are typically $5-10\%$ more stringent.  For the $ttZ$ axial vector 
coupling we observe a somewhat larger variation.  The same statement 
also holds for the sensitivity of the bounds on the normalization
uncertainty 
$\Delta{\cal N}$.  This is illustrated in Fig.~\ref{fig:fig9}, where 
we show $68.3\%$ CL limits for $\Delta F^Z_{1A}$ versus 
$\Delta F^Z_{1V}$ and 300~fb$^{-1}$ at the LHC, using the $\chi^2$ 
test described at the beginning of this section (solid and dashed 
lines), and the log-likelihood method (dotted and dot-dashed lines).  
For both methods, results are shown for $\Delta{\cal N}=30\%$, and 
$\Delta{\cal N}=10\%$. 
The sensitivity bounds on $\Delta F^Z_{1A}$ are seen to vary by as much
as $50\%$ with the statistical method employed, and can be improved by 
as much as factor~2 if $\Delta{\cal N}$ can be reduced from $30\%$ to 
$10\%$.  As in the $t\bar{t}\gamma$ case, a $10\%$ normalization 
uncertainty may be realistic once the NLO QCD corrections to $t\bar{t}Z$ 
production are known.  
 
\subsection{Discussion}
\label{sec:sec5c}

It is instructive to compare the bounds for anomalous $ttV$ couplings
achievable at hadron colliders with the indirect limits from LEP data
and $b\to s\gamma$ decays, and with those projected for a future
$e^+e^-$ linear collider.  The $tt\gamma$ vector and axial vector
couplings are unconstrained by LEP and $b\to s\gamma$ data.  Thus, the 
Tevatron offers a first opportunity to probe these couplings, although
the sensitivity is severely limited by statistics and the
``background'' from initial state radiation.  A much more precise
measurement can be performed at the LHC, which will also be able to
determine the dipole form factors $F^\gamma_{2V,A}$ at the $10-20\%$
level.  Comparing the bounds on $F^\gamma_{2V}$ (Table~\ref{tab:tab2})
with the indirect limits derived in Sec.~\ref{sec:sec2c}, one observes
that the LHC (SLHC) can improve the current bound from $b\to s\gamma$
decays by a factor of about 2~(5).  On the other hand, the limits on
$F^\gamma_{2A}$ which one expects at the LHC are at least one order of
magnitude more stringent than those from $b\to s\gamma$, if one
assumes $F^\gamma_{2A}$ to be real.

The $\epsilon_1$ and $\epsilon_b$ parameters constrain the $ttZ$
vector and axial vector couplings to within a few percent of their SM
values if one assumes that no other source of new physics contributes
to these parameters.  Table~\ref{tab:tab3} and Fig.~\ref{fig:fig8}a
show that it will be impossible to match that precision at the LHC,
even for an integrated luminosity of 3000~fb$^{-1}$.  In contrast,
$\epsilon_2$ and $\epsilon_3$ only constrain a linear combination of
$F^Z_{2V}$ and $F^\gamma_{2V}$.  Thus, $t\bar{t}Z$ production at the
LHC will provide valuable information on the dimension-five $ttZ$
couplings.

\begin{table}
\caption{Sensitivities achievable at $68.3\%$ CL for the anomalous 
$ttV$ ($V=\gamma,\,Z$) couplings $\widetilde F^V_{1V,A}$ and
$\widetilde F^V_{2V,A}$ of Eq.~(\ref{eq:gordon}) at the LHC for
integrated luminosities of 300~fb$^{-1}$, and at an $e^+e^-$ linear
collider operating at $\sqrt{s}=500$~GeV (taken from
Ref.~\protect\cite{Abe:2001nq}).  Only one coupling at a time is
allowed to deviate from its SM value. }
\label{tab:tab4}
\vspace{2mm}
\begin{tabular}{ccc}
coupling & LHC, 300~fb$^{-1}$ & $e^+e^-$~\cite{Abe:2001nq}\\
\tableline 
$\Delta\widetilde F^\gamma_{1V}$ & $\begin{matrix} +0.043 \\[-4pt]
-0.041\end{matrix}$ & $\begin{matrix} +0.047 \\[-4pt]
-0.047\end{matrix}$ , 200~fb$^{-1}$ \\
$\Delta\widetilde F^\gamma_{1A}$ & $\begin{matrix} +0.051 \\[-4pt]
-0.048\end{matrix}$ & $\begin{matrix} +0.011 \\[-4pt]
-0.011\end{matrix}$ , 100~fb$^{-1}$  \\
$\Delta\widetilde F^\gamma_{2V}$ & $\begin{matrix} +0.038 \\[-4pt]
-0.035\end{matrix}$ & $\begin{matrix} +0.038 \\[-4pt]
-0.038\end{matrix}$ , 200~fb$^{-1}$  \\
$\Delta\widetilde F^\gamma_{2A}$ & $\begin{matrix} +0.16 \\[-4pt]
-0.17\end{matrix}$ & $\begin{matrix} +0.014 \\[-4pt]
-0.014\end{matrix}$ , 100~fb$^{-1}$   \\     
\tableline 
$\Delta\widetilde F^Z_{1V}$ & $\begin{matrix} +0.43 \\[-4pt]
-0.83\end{matrix}$ & $\begin{matrix} +0.012 \\[-4pt]
-0.012\end{matrix}$ , 200~fb$^{-1}$  \\
$\Delta\widetilde F^Z_{1A}$ & $\begin{matrix} +0.14 \\[-4pt]
-0.14\end{matrix}$ & $\begin{matrix} +0.013 \\[-4pt]
-0.013\end{matrix}$ , 100~fb$^{-1}$  \\
$\Delta\widetilde F^Z_{2V}$ & $\begin{matrix} +0.38 \\[-4pt]
-0.50\end{matrix}$ & $\begin{matrix} +0.009 \\[-4pt]
-0.009\end{matrix}$ , 200~fb$^{-1}$  \\
$\Delta\widetilde F^Z_{2A}$ & $\begin{matrix} +0.50 \\[-4pt]
-0.51\end{matrix}$ & $\begin{matrix} +0.052 \\[-4pt]
-0.052\end{matrix}$ , 100~fb$^{-1}$     
\end{tabular}
\end{table}

The most complete study of $t\bar{t}$ production at a future $e^+e^-$
linear collider for general $ttV$ ($V=\gamma,\,Z$) couplings so far is
that of Ref.~\cite{Abe:2001nq}.  It uses the parameterization of
Eq.~(\ref{eq:gordon}) for the $ttV$ vertex function.  In order to
compare the bounds of Ref.~\cite{Abe:2001nq} with those anticipated
at the LHC, the limits derived in Secs.~\ref{sec:sec5a}
and~\ref{sec:sec5b} have to be converted into bounds on $\widetilde
F^V_{1V,A}$ and $\widetilde F^V_{2V,A}$ (see
Eqs.~(\ref{eq:rel1}--\ref{eq:rel4})).  Table~\ref{tab:tab4} compares
the bounds we obtain for $\widetilde F^V_{1V,A}$ and $\widetilde
F^V_{2V,A}$ with those reported in Ref.~\cite{Abe:2001nq} for an
$e^+e^-$ linear collider operating at $\sqrt{s}=500$~GeV, which
assumes a linear polarization of ${\cal P}^-={\cal P}^+=0.8$ for both
electron and positron beams.  Ref.~\cite{Abe:2001nq} lists sensitivity
bounds only for the case that only one coupling at a time is allowed
to deviate from its SM value, as we do for the LHC in
Table~\ref{tab:tab4}.  Furthermore, we show limits only for an
integrated luminosity of 300~fb$^{-1}$.  The results of
Table~\ref{tab:tab4} demonstrate that a linear collider, with the 
exception of $F^\gamma_{1V}$ and $F^\gamma_{2V}$, will be able to 
considerably improve the sensitivity limits which can be achieved at 
the LHC, in particular for the $ttZ$ couplings. For the SLHC, with 
3000~fb$^{-1}$, we obtain bounds for the anomalous $ttV$ couplings 
which are a factor~$2-3$ more stringent than those shown in 
Table~\ref{tab:tab4}.  Thus, even if the SLHC operates first, a linear 
collider will still be able to improve the $ttZ$ anomalous coupling 
limits by at least a factor~3.  It should be noted, however, that this 
picture could change once correlations between different non-standard 
$ttV$ couplings are taken into account.  Unfortunately, so far, no 
realistic studies for $e^+e^-\to t\bar{t}$ which include these 
correlations have been performed.  We found that there are significant 
correlations between the various $ttV$ couplings at the LHC.  Since 
both $tt\gamma$ and $ttZ$ contribute to $e^+e^-\to t\bar{t}$, the 
correlations may even be larger at a $e^+e^-$ linear collider. More 
detailed studies are needed in order to answer this question.

Our calculation of sensitivity bounds is subject to several
uncertainties.  The cross sections of the main backgrounds,
$t\bar{t}j$ and $Zb\bar{b}+4j$ production, are proportional to
$\alpha_s^3$ and $\alpha_s^6$, respectively, whereas the signal
cross section scales as $\alpha_s^2$.  The background thus depends
more strongly on the factorization and renormalization scale than the
signal. The background normalization can be fixed by relaxing the
$t\bar{t}\gamma$ (Eqs.~(\ref{eq:cuts1}--\ref{eq:cuts3})) or
$t\bar{t}Z$ selection cuts (Eqs.~(\ref{eq:cuts1}), (\ref{eq:cuts5}) and
(\ref{eq:cuts4} - \ref{eq:cuts7})), measuring the cross section in that
background-dominated region of phase space, and then extrapolating
back to the analysis region.  Since $t\bar{t}j$ production is the
dominant source of background in the $t\bar{t}\gamma$ case, S:B
sensitively depends on the jet photon misidentification probability,
$P_{j\to\gamma}$.  This has been measured at the Tevatron, at least for
small values of the photon transverse momentum.  For the LHC, we have
relied on ATLAS and CMS simulations.  Finally, in calculating limits
we have ignored the background from $W\gamma+$jets and $WZ+$~jets
production, where two of the jets are misidentified as $b$-quarks.
While these backgrounds should be very small at the Tevatron and LHC,
they may be more important at the SLHC.  Fortunately, the total
background for both $t\bar{t}\gamma$ and $t\bar{t}Z$ production is
relatively small and hardly affects the ultimate sensitivity limits.
Increasing the background cross section by a factor~2, for example,
weakens the bounds by only a few percent.

In our analysis, we have assumed that both $b$ quarks are tagged.  If
events with only one $b$ tag can be utilized, the sensitivity bounds
can be improved by up to a factor~1.5.  However, detailed background
calculations are needed before a firm conclusion can be drawn.  The
same statement applies to the $\sla{p}_Tb\bar{b}+4j$ final state,
which has the potential of improving the sensitivity limits for
anomalous $ttZ$ couplings by as much as a factor~1.7.  Finally, we
stress that our calculation was based on a simple $\chi^2$ test.  More
powerful statistical tools such as those used in the recent
re-analysis of the top quark mass~\cite{topmass}, or a neural net
analysis, may further improve the limits.

\section{Summary and Conclusions}
\label{sec:sec6}

Currently, little is known about top quark couplings to the photon and
$Z$ boson.  There are no direct measurements of these couplings;
indirect measurements, using LEP data, tightly constrain only the
$ttZ$ vector and axial vector couplings.  All others are only very
weakly constrained by LEP and/or $b\to s\gamma$ data.  The $ttV$
($V=\gamma,\,Z$) couplings can be measured directly in $e^+e^-\to
t\bar{t}$ at a future $e^+e^-$ linear collider. However, such a
machine is at least a decade away. In addition, the process $e^+e^-\to
t\bar{t}$ is simultaneously sensitive to $tt\gamma$ and $ttZ$
couplings, and significant cancellations between various couplings may
occur.

In this paper, we have considered $t\bar{t}\gamma$ production
(including radiative top decay, $t\to Wb\gamma$, in $t\bar{t}$ events)
and $t\bar{t}Z$ production at hadron colliders as tools to measure the
$ttV$ couplings.  We calculated the signal cross sections, taking into
account the full set of contributing Feynman diagrams.  In
$t\bar{t}\gamma$ production, we concentrated on the $\gamma\ell\nu
b\bar{b}jj$ final state.  For $t\bar{t}Z$ production, we assumed that
the $Z$ boson decays leptonically, $Z\to{\ell'}^+{\ell'}^-$, and
investigated the ${\ell'}^+{\ell'}^-\ell\nu b\bar{b}jj$ (trilepton)
and ${\ell'}^+{\ell'}^-b\bar{b}+4j$ (dilepton) final states.  All
relevant background processes were included.  Once $t\bar{t}\gamma$ or
$t\bar{t}Z$ selection cuts are imposed, the total background is
substantially smaller than the signal.  The dominant background source
for $t\bar{t}\gamma$ events is QCD $t\bar{t}j$ production, where one
jet is misidentified as a photon.  For $t\bar{t}Z$ production,
$Zb\bar{b}+4j$ production and singly-resonant processes are the
main sources.  In all our calculations we assumed that both $b$ quarks
are tagged.

At the Tevatron, the $t\bar{t}Z$ cross section is too small to be
observable. The $t\bar{t}\gamma$ cross section is large enough to
allow for a first, albeit not very precise, test of the $tt\gamma$
vector and axial vector couplings, provided that an integrated
luminosity of more than 5~fb$^{-1}$ can be accumulated. No useful
limits on the dipole form factors $F^\gamma_{2V,A}$ can be
obtained. Since $q\bar{q}$ annihilation dominates at Tevatron energies,
initial state photon radiation severely limits the sensitivity of
$t\bar{t}\gamma$ production to anomalous top quark couplings.

This is not the case at the LHC where gluon fusion is the dominant
production mechanism.  Combined with a much larger cross section, this
results in much-improved sensitivity limits.  Already with an
integrated luminosity of 30~fb$^{-1}$, which is expected after the
first 3~years of operation, one can probe the $tt\gamma$ couplings
with a precision of about $10-35\%$ per experiment.  With
300~fb$^{-1}$, which corresponds to 3~years of running at design
luminosity, a $4-7\%$ measurement of the $tt\gamma$ vector and axial
vector couplings can be expected, while the dipole form factors
$F^\gamma_{2V,A}$ can be measured with $20\%$ accuracy.  Finally, if
the luminosity of the LHC can be upgraded by a factor of~10 (the SLHC
program) without significant loss of particle detection efficiency for
photons, leptons and $b$ quarks, these limits can be improved by
another factor $2-3$.

The $t\bar{t}Z$ cross section with leptonic $Z$ decays is roughly a
factor~20 smaller than the $t\bar{t}\gamma$ rate.  It is therefore not
surprising that the sensitivity limits on the $ttZ$ couplings are
significantly weaker than those which one expects for the $tt\gamma$
couplings.  We found that, for 300~fb$^{-1}$, the $ttZ$ vector (axial
vector) couplings can be measured with a precision of $45-85\%$
($15-20\%$), and $F^Z_{2V,A}$ with a precision of $50-55\%$.  At the
SLHC, these bounds can be improved by factors of $1.4-2$ ($\approx 3$)
and~1.6, respectively.

In our analysis, we conservatively assumed that both $b$ quarks are
tagged, and used a simple $\chi^2$ test to derive sensitivity limits.
If single-$b$-tag events can be utilized, the sensitivity bounds can
be significantly strengthened.  Further improvements could also result
from using more powerful statistical tools, similar to those which
have been used recently to measure the top quark mass~\cite{topmass}.

\acknowledgements

We would like to thank E.~Boos, R.~Demina, J.~Parsons, P.~Tipton and
D.~Toback for useful discussions.  One of us (U.B.) would like to
thank the Fermilab Theory Group, where part of this work was carried
out, for its generous hospitality.  This research was supported in
part by the National Science Foundation under grant No.~PHY-0139953
and the Department of Energy under grant DE-FG02-91ER40685.  Fermilab
is operated by Universities Research Association Inc. under Contract
No. DE-AC02-76CH03000 with the U.S. Department of Energy.




\end{document}